\documentclass[letterpaper]{article}

\usepackage[utf8]{inputenc}
\usepackage[T1]{fontenc}
\usepackage{geometry}
\usepackage{setspace}

\usepackage[journal=jctcce,articletitle=true]{achemso}

\usepackage{amsmath}
\usepackage{amssymb}
\usepackage{amsfonts}
\usepackage{array}
\usepackage{booktabs}
\usepackage{bm}
\usepackage{caption}
\usepackage{etoolbox}
\usepackage{flafter}
\usepackage{float}
\usepackage{graphicx}
\usepackage{longtable}
\usepackage{microtype}
\usepackage{multirow}
\usepackage{needspace}
\usepackage[section]{placeins}
\usepackage{siunitx}
\usepackage{subcaption}
\usepackage{threeparttable}
\usepackage{url}
\usepackage{xcolor}
\usepackage{hyperref}
\usepackage{tabularx}
\usepackage{ragged2e}
\hypersetup{
  hidelinks,
  pdftitle={DeepHartree: A Poisson-Coupled Neural Field for One-Shot Density Functional Theory},
  pdfauthor={Jiankun Wu; Jinming Fan; Chao Qian; Shaodong Zhou}
}

\newcolumntype{Y}{>{\RaggedRight\arraybackslash}X}
\newcolumntype{L}[1]{>{\RaggedRight\arraybackslash}p{#1}}

\usepackage{authblk}
\author[1]{Jiankun Wu}
\author[1]{Jinming Fan}
\author[1,2]{Chao Qian}
\author[1,2]{Shaodong Zhou*}
\affil[1]{College of Chemical and Biological Engineering, Zhejiang University, Hangzhou, CN 310027}
\affil[2]{Institute of Zhejiang University--Quzhou, 324000 Quzhou (P.R. China)}

\title{DeepHartree: A Poisson-Coupled Neural Field for One-Shot Density Functional Theory}
\date{*Corresponding author. Email: szhou@zju.edu.cn\\
Other author emails: jiankun.wu@zju.edu.cn; fanjinming@zju.edu.cn; qianchao@zju.edu.cn}

\begin{document}

\raggedbottom
\maketitle

\begin{abstract}
Linear-combination-of-atomic-orbital (LCAO) density functional theory (DFT) incurs steep costs when it constructs Coulomb terms and iterates the self-consistent field (SCF) equations. Matrix-learning approaches can bypass parts of this workflow, but their outputs inherit the dimensions and conventions of a fixed orbital basis. We introduce DeepHartree, a Poisson-coupled neural field that connects continuous real-space prediction to LCAO electronic structure. An E(3)-equivariant network predicts the Hartree potential, and the Poisson equation converts this potential into electron density. Atom-centred Gaussian fields resolve the near-nuclear region, while a molecule-level correction enforces the electron count. {Analytical Gaussian integrals and numerical quadrature connect the predicted fields to the Kohn--Sham matrix.} {One diagonalization provides estimates of frontier orbital energies and the occupied subspace, and yields the density matrix used to evaluate energy components and initialize SCF.}
{Mean molecular weighted normalized mean absolute errors (wNMAEs) are 0.110\% for QM9 and 0.199\% for VQM24 on PySCF level-0 grids, increasing to 1.366\% and 2.218\%, respectively, on level-3 grids. Models trained on def2-SVP densities transfer without retraining to def2-TZVP and def2-QZVP, retaining total-energy accuracy and occupied-subspace agreement.}
{The Hybrid method combines finite differences and first-order automatic differentiation with analytical Gaussian integrals and multilevel quadrature to accelerate field-to-matrix conversion. It achieves near-linear density-matrix construction in both alkanes and PEG chains. For the 1004-atom PEG chain, this construction is 8.75-fold faster than full AD and 7.34-fold faster than converged GPU4PySCF SCF.}
Without fine-tuning, the QM9 model attains a total-energy MAE of {9.437}~meV atom$^{-1}$ on 1,000 larger OE62 molecules.
{Across the three basis sets, DeepHartree initialization reduces mean QM9 SCF iteration counts by 15.3--15.6\%. For a 620-atom organic cage, it reduces total SCF wall-clock time by 12.9\%, including initialization.}
These results establish continuous electrostatic fields as an accurate and scalable interface between machine learning and LCAO DFT.
\end{abstract}

% !TEX root = ../main.tex
\section{Introduction}

Ab initio electronic-structure calculations provide the energies, forces, and electronic observables used throughout computational chemistry. Their cost, however, grows steeply with system size and basis-set expansion. This scaling limits routine calculations for large molecular systems and repeated simulations.

Machine learning (ML) methods pursue two main routes around this cost. The first predicts post-SCF properties, including energies and forces. ML potentials \citep{batzner20223,yu2025extending,cui2024geometry,musaelian2023learning,batatia2206mace,unke2019physnet,wang2024enhancing} bypass the iterative SCF calculation and support efficient molecular simulation. Their scalar outputs omit the intermediate electronic structure needed to reconstruct general quantum-chemical observables. The second route predicts information-rich variables inside an electronic-structure calculation. Early work learned local atomic-orbital DFT Hamiltonian elements for band-structure and transport calculations \citep{hegde2017machine}. For molecular systems, \citet{schutt2019unifying} predicted the Hamiltonian matrix, and \citet{unke2021se} introduced an equivariant formulation that improved its accuracy. More recent models impose Hamiltonian self-consistency through deep equilibrium networks \citep{wang2024infusing} or learn the one-electron reduced density matrix to recover energies, forces, orbitals, and one-electron observables \citep{shao2023machine}. Matrix models remain tied to their orbital representation: a change of basis alters both the matrix dimension and its numerical entries, which complicates transfer between basis sets and software conventions.

Other intermediate representations retain complementary electronic-structure information. DOS-based models have learned either low-dimensional patterns of the total density of states \citep{yeo2019pattern} or grid-resolved charge densities and local densities of states (LDOS) \citep{chandrasekaran2019solving}. Building on the LDOS formulation, the MALA framework reconstructed finite-temperature DFT observables for solid and liquid aluminum \citep{ellis2021accelerating}. It also transferred a model trained on small cells to systems containing more than $10^5$ atoms \citep{fiedler2023predicting} and covered both ionic and electronic temperatures \citep{fiedler2023temperature}. These studies show that field and spectral representations retain access to several downstream observables while exposing intermediate electronic information.

The Hohenberg-Kohn theorems place electron density at the center of ground-state DFT \citep{hohenberg1964inhomogeneous}. They establish that the density $\rho(r)$ uniquely determines the ground state of an interacting many-particle system. This result has motivated ML models centered on density. \citet{brockherde2017bypassing} learned the Hohenberg-Kohn map from an external potential to the ground-state density and evaluated the corresponding energy. Local-environment models reconstructed ground-state densities in solids \citep{schmidt2018machine}. In a distinct property-learning approach, multiscale wavelet representations of molecular structure supported molecular-property prediction \citep{eickenberg2018solid}. Density-based $\Delta$-learning corrected baseline DFT energies and forces toward higher-level references \citep{dick2019machine}. \citet{lentz2020predicting} mapped PBE charge densities to HSE band gaps, and physics-informed architectures embedded orbital and density generation into end-to-end property learning \citep{tsubaki2020quantum}. More recently, \citet{rackers2023recipe} demonstrated transfer from small water clusters to systems beyond the sizes accessible to their reference calculations.

Within direct real-space density prediction, \citet{jorgensen2022equivariant} used PaiNN \citep{schutt2021equivariant} to predict electron densities on grids. Higher-order tensor representations further improved density accuracy \citep{koker2024higher}. Using a different strategy, \citet{li2025image} combined ResNet and super-resolution methods to reconstruct densities from a superposition of atomic densities (SAD). These grid-based studies commonly use plane-wave (PW) basis sets and pseudopotentials. Pseudopotentials remove sharp core cusps from the target density, producing smoother scalar fields that integrate directly with PW software. FFT-based solvers \citep{cooley1965algorithm,frigo2005design} underpin plane-wave codes such as VASP \citep{kresse1996efficient}, providing a natural route for transferring grid fields into PW workflows.

Researchers have also applied ML at specific numerical stages of electronic-structure calculations. \citet{meyer2020machine} jointly learned an orbital-free kinetic-energy functional and its derivative to stabilize variational density optimization. \citet{ku2019machine} used ML to select collocation points for the Kohn-Sham equation. Extensive neural networks enforce additive scaling through spatial domain decomposition \citep{mills2019extensive}. At the level of derived electronic information, \citet{ferreira2020machine} mapped local structural descriptors to quantum-mechanical bond indicators in metallic glasses. Graph neural networks have also predicted molecular X-ray absorption spectra \citep{carbone2020transferable}. These methods target numerical acceleration, scalar corrections, spectra, matrices, or fields. Each target provides a different interface to a downstream quantum-chemistry workflow.

The remaining gap lies at the interface between density learning and LCAO quantum chemistry. Many density models target periodic or PW calculations, and most studies evaluate the predicted density itself. Fewer studies convert a learned all-electron field into a Kohn--Sham matrix for molecular LCAO calculations or use it to recover a broader electronic spectrum. This interface is difficult because AO codes assemble the Kohn--Sham matrix from several basis-dependent terms instead of accepting a real-space density directly. A matrix predictor also inherits the dimensions and conventions of the AO basis used during training.

Recent work by \citet{liutowards} narrows this interface gap by predicting electron-density coefficients in a compact auxiliary basis. Their representation converts directly into an SCF initial guess and demonstrates transfer across system size, AO basis, and exchange--correlation functional. A model trained on molecules with at most 20 atoms accelerates systems containing as many as 900 atoms. The auxiliary expansion reduces the output size and integrates cleanly with density-fitting implementations. Its expressiveness controls the density-fitting error \citep{weigend2006accurate}. {DeepHartree takes a complementary representation-level approach by parameterizing a continuous Hartree potential and deriving density through the Poisson equation. Analytical Gaussian integrals and numerical quadrature map the resulting fields into the target LCAO basis.}

DeepHartree implements this route as a Poisson-coupled neural field (PCNF) (Fig.~\ref{fig:framework}). An E(3)-equivariant network predicts the Hartree potential, and the Poisson equation yields the electron density. Delta-learned Gaussian fields resolve the near-nuclear region, and a molecule-level correction enforces the electron count.{The same trained field can be mapped into different target AO bases without changing the network parameters or requiring an AO-dimension-specific output head. Diagonalizing the resulting Kohn--Sham matrix yields one-shot electronic properties and an SCF initial density matrix, while the local field representation supports size extrapolation.}

DeepHartree makes four contributions:
\begin{itemize}
	\item \textbf{Field-to-LCAO architecture.} {DeepHartree couples an equivariant Hartree-potential model to the Poisson equation, a molecular charge constraint, and analytical and numerical LCAO projection.} This design preserves a defined potential--density relation and avoids an output layer tied to a fixed AO matrix dimension.
	
	\item {\textbf{Density accuracy and cross-basis transfer.} Evaluation on 39,214 test molecules establishes density and one-shot electronic-structure accuracy across QM9 and VQM24. Models trained on def2-SVP densities transfer directly to def2-TZVP and def2-QZVP without retraining, preserving total-energy accuracy and occupied-subspace agreement.}
	
	\item {\textbf{Efficient hybrid field-to-matrix conversion.} Hybrid combines seven-point finite differences and first-order automatic differentiation with analytical Gaussian Coulomb integrals and multilevel exchange--correlation quadrature. It retains total-energy accuracy across the tested bases and enables near-linear density-matrix construction for systems containing up to 1004 atoms.}

	\item {\textbf{Size transfer and SCF acceleration.} Without fine-tuning, AD evaluation of the QM9-trained model achieves a total-energy MAE of 9.437~meV atom$^{-1}$ on 1000 larger OE62 molecules \citep{stuke2020atomic}. Model-derived density matrices reduce mean QM9 SCF iteration counts across all three basis sets. For a 620-atom organic cage, model-based initialization lowers total SCF wall-clock time by 12.9\%, including the cost of constructing the initial guess.}
\end{itemize}

% !TEX root = ../main.tex
\section{Results}

\subsection{Overview of DeepHartree}

DeepHartree connects a continuous neural field to an LCAO electronic-structure calculation (Fig.~\ref{fig:framework}). An equivariant graph network reads the molecular geometry and predicts short-range grid corrections together with atom-centred long-range terms of the Hartree potential. Fixed atom-centred Gaussian terms represent the near-nuclear density. Their sum gives $V_H(\mathbf r)$, from which automatic differentiation obtains the electron density and the gradient required by a GGA functional,

\begin{equation}
	\rho(\mathbf r)=-\frac{1}{4\pi}\nabla^2V_H(\mathbf r), \qquad
	E_{\mathrm{xc}}^{\mathrm{GGA}}[\rho]=\int \varepsilon_{\mathrm{xc}}\!\left(\rho(\mathbf r),\nabla\rho(\mathbf r)\right)\,\mathrm d^3\mathbf r.
\end{equation}

Numerical integration maps $V_H$, $\rho$, and $\nabla\rho$ to the Coulomb and exchange--correlation matrices. Combining them with the analytical kinetic and nuclear-attraction matrices yields a Kohn--Sham matrix. One generalized eigendecomposition then produces orbitals, eigenvalues, and a closed-shell AO density matrix for non-self-consistent property evaluation.

The revised output layer enforces the electron count through a molecule-level correction. Let $q_{ik}$ be a predicted coefficient, $N_{\mathrm{func}}$ be the total number of learnable Gaussian shells in the molecule, and $Q_{\mathrm{fixed}}$ be the sum of fixed Gaussian coefficients. Both sums below run over these same $N_{\mathrm{func}}$ learnable shells. We add an identical charge correction to every predicted coefficient,

\begin{equation}
	q'_{ik}=q_{ik}+
	\frac{N_{\mathrm e}-Q_{\mathrm{fixed}}-\sum_{jl}q_{jl}}
	{N_{\mathrm{func}}},
	\qquad Q_{\mathrm{fixed}}+\sum_{ik}q'_{ik}=N_{\mathrm e}.
\end{equation}

\Needspace{12\baselineskip}
To see why this coefficient constraint conserves charge, we define the total far-field coefficient as $Q_{\infty}=Q_{\mathrm{fixed}}+\sum_{ik}q'_{ik}$. Because the grid-readout correction decays to zero at its cutoff, the atom-centred terms give $V_H(\mathbf r)=Q_{\infty}/r+\mathcal O(r^{-2})$ as $r\to\infty$. Integrating the Poisson relation over all space then gives

\begin{align}
	\int_{\mathbb R^3}\rho(\mathbf r)\,\mathrm d^3\mathbf r
	&=-\frac{1}{4\pi}\int_{\mathbb R^3}\nabla^2V_H(\mathbf r)\,\mathrm d^3\mathbf r \\
	&=-\frac{1}{4\pi}\lim_{R\to\infty}\oint_{S_R}\nabla V_H(\mathbf r)\cdot\mathrm d\mathbf S \\
	&=Q_{\infty}=Q_{\mathrm{fixed}}+\sum_{ik}q'_{ik}=N_{\mathrm e}.
\end{align}

The second equality is Gauss's divergence theorem. Thus, correcting the molecular sum of the learnable coefficients fixes the integrated electron number of each neutral closed-shell molecule.

\begin{figure}[!t]
	\centering
	\includegraphics[width=\linewidth]{img/framework.pdf}
	\caption{\textbf{DeepHartree maps a continuous Hartree potential to an LCAO density matrix.} An equivariant network predicts learnable potential terms, the Poisson equation yields the density, and numerical integration assembles the Kohn--Sham matrix. Its diagonalization yields one-shot electronic properties and the AO density matrix.}
	\label{fig:framework}
\end{figure}

\subsection{Density accuracy across QM9 and VQM24}

We evaluated the PaiNN backbone \citep{schutt2021equivariant} on every molecule in both fixed test splits. \noindent{For each molecule, the $L_1$ and $L_2$ norms summarize the pointwise density error.} We define the weighted normalized mean absolute error (wNMAE) from the continuous density integral and evaluate it using the molecular quadrature grid,

\begin{equation}
\begin{aligned}
\mathrm{wNMAE}_i
&=
\frac{
\displaystyle\int
\left|\hat{\rho}_i(\mathbf r)-\rho_i(\mathbf r)\right|
\,\mathrm d\mathbf r
}{
\displaystyle\int
\left|\rho_i(\mathbf r)\right|
\,\mathrm d\mathbf r
}
=
\frac{
\displaystyle\sum_p w_{ip}
\left|\hat{\rho}_{ip}-\rho_{ip}\right|
}{
\displaystyle\sum_p w_{ip}
\left|\rho_{ip}\right|
},
\\[3pt]
{ L_{1,i}}
&{
=
\int
\left|\hat{\rho}_i(\mathbf r)-\rho_i(\mathbf r)\right|
\,\mathrm d\mathbf r
=
\sum_p w_{ip}
\left|\hat{\rho}_{ip}-\rho_{ip}\right|
},
\\[3pt]
{ L_{2,i}}
&{
=
\sqrt{
\int
\left|\hat{\rho}_i(\mathbf r)-\rho_i(\mathbf r)\right|^2
\,\mathrm d\mathbf r
}
=
\sqrt{
\sum_p w_{ip}
\left|\hat{\rho}_{ip}-\rho_{ip}\right|^2
}
}.
\end{aligned}
\label{eq:wnmae}
\end{equation}
{Here, $p$ indexes grid points and $w_{ip}$ denotes the signed PySCF quadrature weight for molecule $i$. All reported density integrals retain the signs of these weights.}

\begin{table}[!htbp]
    \centering
    \caption{\noindent{\textbf{Molecular density errors and charge diagnostics for QM9 and VQM24.} Values are means over molecules at PySCF grid levels 0 and 3, evaluated with signed quadrature weights. For each molecule, $Q_-=\sum_p w_p\max(-\hat\rho_p,0)$, $f_{Q_-}=Q_-/\sum_p w_p|\hat\rho_p|$, and $\Delta N=|\sum_p w_p(\hat\rho_p-\rho_p)|$.}}
    \label{tab:density_errors}
    \resizebox{\linewidth}{!}{%
	{
    \begin{tabular}{lrrrrrr}
        \toprule
        Dataset & wNMAE (\%) & $L_1$ (e) & $L_2$ ($\mathrm{e}\,\mathrm{Bohr}^{-3/2}$) & $Q_-$ ($10^{-3}\,\mathrm{e}$) & $f_{Q_-}$ ($10^{-3}\%$) & $\Delta N$ ($10^{-3}\,\mathrm{e}$) \\
        \midrule
        \multicolumn{7}{l}{\textbf{Level 0}} \\
		\midrule
        QM9   & 0.110 & 0.072 & 0.025 & 0.197  & 0.304  & 3.763 \\
        VQM24 & 0.199 & 0.125 & 0.116 & 0.104  & 0.186  & 8.583 \\
        \midrule
        \multicolumn{7}{l}{\textbf{Level 3}} \\
		\midrule
        QM9   & 1.366 & 0.899 & 0.354 & 21.625 & 33.199 & 16.053 \\
        VQM24 & 2.218 & 1.431 & 0.869 & 12.810 & 20.357 & 18.665 \\
        \bottomrule
    \end{tabular}%
	}
    }
\end{table}

\begin{figure}[!t]
    \centering
    \includegraphics[width=\linewidth]{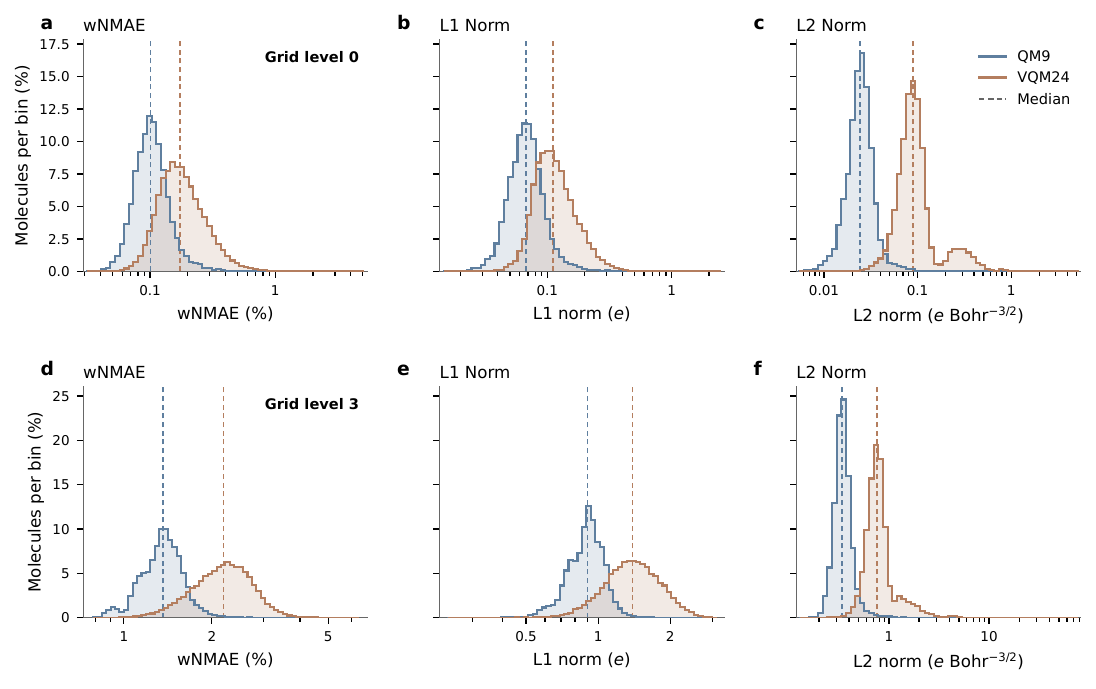}
    \caption{\noindent{\textbf{Molecule-level density-error distributions at PySCF grid levels 0 and 3.} Panels \textbf{a}--\textbf{c} show the distributions at grid level 0, whereas panels \textbf{d}--\textbf{f} show those at grid level 3. Within each row, the panels report wNMAE, the $L_1$ norm, and the $L_2$ norm, respectively, evaluated with signed quadrature weights. The histograms compare QM9 ($n=13{,}389$ molecules) with VQM24 ($n=25{,}825$ molecules). Each histogram is independently normalized as the percentage of molecules per bin. Dashed vertical lines indicate the median of each distribution.}}
    \label{fig:level0_level3_distributions}
\end{figure}

\par\begingroup
DeepHartree achieves high density accuracy on the level-0 grid, with mean wNMAEs of 0.110\% for QM9 and 0.199\% for VQM24 (Table~\ref{tab:density_errors}). The small $L_1$ and $L_2$ errors further demonstrate close agreement with the reference densities on this grid. The distributions in Fig.~\ref{fig:level0_level3_distributions} show higher median errors for VQM24 across all three metrics at both grid levels. Thus, the difference extends to typical molecules rather than arising solely from isolated outliers, consistent with VQM24's greater elemental and structural diversity.

Grid refinement reveals errors that are less apparent in the coarse-grid evaluation. On the level-3 grid, mean wNMAE reaches approximately 1.37\% for QM9 and 2.22\% for VQM24 with the same checkpoints and molecules. Both integral norms also increase, so the change extends beyond the normalization of wNMAE. These results demonstrate sensitivity to spatial sampling and signed quadrature, with finer-grid evaluation revealing larger density discrepancies than coarse-grid evaluation.

Across both datasets and grid levels, mean $Q_-$ and $f_{Q_-}$ remain below 0.022~e and 0.034\%, respectively, indicating small integrated negative-density contributions. The electron-count MAE relative to reference densities integrated on the same grid is 0.0038~e for QM9 and 0.0086~e for VQM24 at level 0, and 0.0161~e and 0.0187~e, respectively, at level 3 (Table~\ref{tab:density_errors}). These small differences support accurate total-charge recovery under numerical quadrature, consistent with the model's analytical charge constraint.
\par\endgroup

\begin{table*}[!htbp]
    \centering
    \caption{\noindent{\textbf{Mean molecule-level wNMAE for molecules containing selected heteroatoms.}
    Values report the arithmetic mean wNMAE, evaluated with signed quadrature weights at PySCF grid level 3, for molecules containing each indicated element. Element-defined subsets may overlap because one molecule can contain multiple heteroatoms. Dashes indicate that the corresponding element is absent from the dataset.}}
    \label{tab:element_wnmae}
    \setlength{\tabcolsep}{7pt}
    {
    \begin{tabular}{lcccccccc}
        \toprule
        Dataset & N & O & F & Si & P & S & Cl & Br \\
        \midrule
        QM9
        & 1.440\%
        & 1.389\%
        & 1.499\%
        & --
        & --
        & --
        & --
        & -- \\
        VQM24
        & 2.308\%
        & 2.324\%
        & 2.129\%
        & 2.052\%
        & 2.173\%
        & 2.346\%
        & 2.122\%
        & 1.704\% \\
        \bottomrule
    \end{tabular}
    }
\end{table*}

\par\begingroup
Density accuracy remains consistent across the heteroatom-containing subsets (Table~\ref{tab:element_wnmae}). Subset means cluster closely within QM9 and span approximately 1.70--2.35\% in VQM24, with Br-containing molecules at the lower end. The broader elemental coverage therefore retains comparable relative accuracy across the represented chemical environments, supporting transferability within the sampled chemical space.
\par\endgroup

\paragraph{{Ablation studies.}}

\par\begingroup
The QM9 level-0 ablations establish the benefits of Gaussian density priors and charge conservation (Table~\ref{tab:qm9_density_ablation}). Fixed atomic Gaussians reduce wNMAE from 0.271\% to 0.119\%, and learnable shells further reduce it to approximately 0.110\%. By representing the sharp near-nuclear density analytically, the fixed Gaussians provide a strong physical prior that compresses the local residual's dynamic range. Learnable shells adapt this baseline to each atom's environment, further simplifying the local density correction. Local-Only and With Fixed Density yield undefined $L_2$ values because PySCF's pruned grids include Lebedev rules with negative weights. Large errors at these points can make $\sum_p w_p|\hat\rho_p-\rho_p|^2$ negative, yielding NaN upon taking its square root despite small $L_1$ and wNMAE. Enforcing charge conservation lowers $L_2$ from approximately 0.037 to 0.025~$\mathrm{e}\,\mathrm{Bohr}^{-3/2}$ while leaving wNMAE and $L_1$ nearly unchanged. This improvement is consistent with better control of large local residuals at a similar overall absolute-error level, combining adaptable Gaussian shells with the correct electron count.
\par\endgroup

\begin{table}[!htbp]
    \centering
    \caption{{\textbf{Density-model ablations on the QM9 test set at PySCF grid level 0.} Values are molecular arithmetic means evaluated with signed quadrature weights. Local-Only contains no Gaussian terms; With Fixed Density adds the fixed atomic prior; Without Charge Conservation also includes learnable shells but omits their molecular charge correction. Full DeepHartree includes all components. NaN denotes an undefined $L_2$ arising from a negative discrete squared-error sum.}}
    \label{tab:qm9_density_ablation}
    {
    \begin{tabular}{lrrr}
        \toprule
        Model & wNMAE (\%) & $L_1$ (e) & $L_2$ ($\mathrm{e}\,\mathrm{Bohr}^{-3/2}$) \\
        \midrule
        Local-Only & 0.271 & 0.177 & NaN \\
        With Fixed Density & 0.119 & 0.079 & NaN \\
        Without Charge Conservation & 0.110 & 0.072 & 0.037 \\
        Full DeepHartree & 0.110 & 0.072 & 0.025 \\
        \bottomrule
    \end{tabular}
    }
\end{table}

\subsection{One-shot quantum-chemical property prediction}

\noindent\paragraph{{Hybrid Method}}

\begin{figure}[!t]
	\centering
	\includegraphics[width=\linewidth]{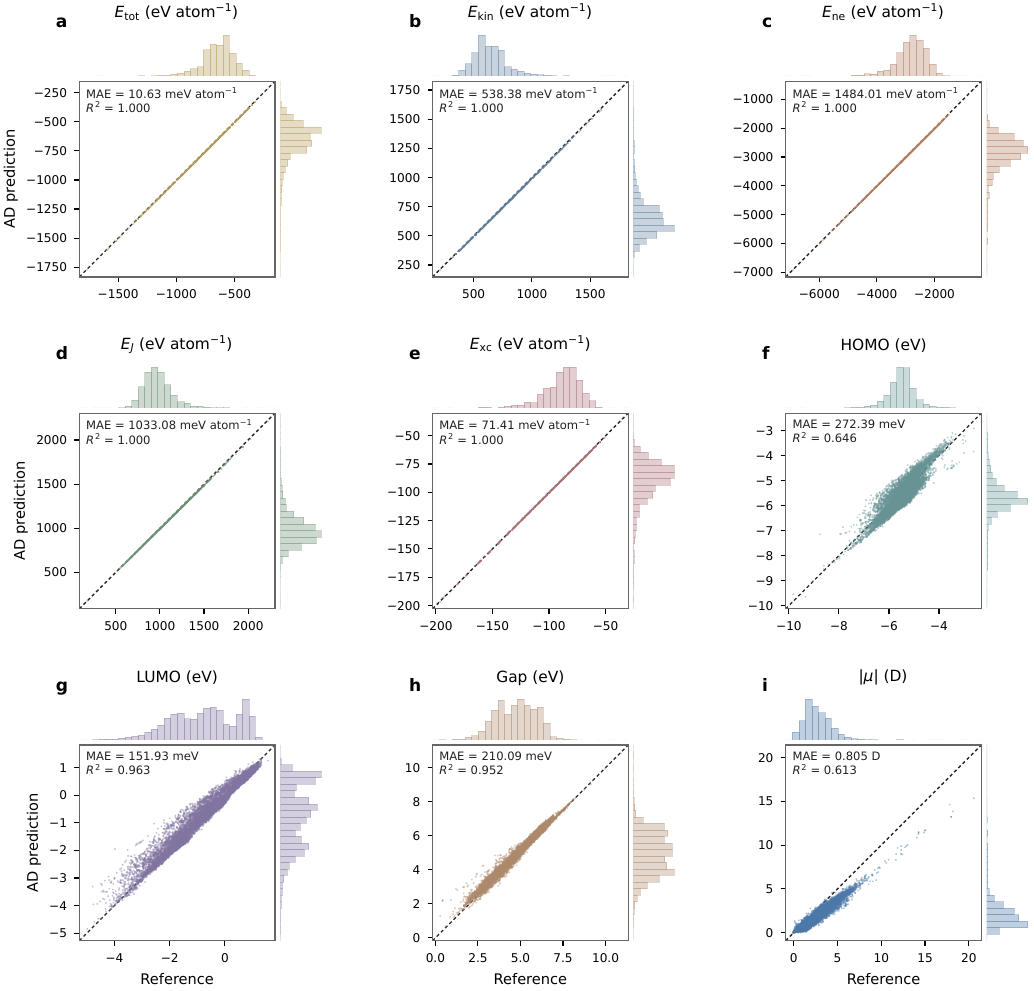}
	\caption{\noindent{\textbf{AD-derived one-shot properties agree with QM9 references.} Predicted and reference values are compared for \textbf{a}, total energy; \textbf{b}--\textbf{e}, kinetic, electron--nuclear, Coulomb, and exchange--correlation energy components; \textbf{f}--\textbf{h}, HOMO, LUMO, and the HOMO--LUMO gap; and \textbf{i}, dipole-moment magnitude.  Dashed lines denote exact agreement. Upper and right marginal histograms show the reference and predicted distributions, respectively. Panel annotations report the mean absolute error (MAE) and coefficient of determination ($R^2$) for all 13,389 QM9 test molecules using def2-SVP.}}
	\label{fig:qm9_ad_properties}
\end{figure}

We constructed a Kohn--Sham matrix from each predicted density and evaluated energy components and frontier orbital energies after one diagonalization (Fig.~\ref{fig:qm9_ad_properties}). The same figure reports dipole-moment magnitudes obtained from the model-derived atomic charges, while the occupied-subspace overlap and fixed-point consistency are summarized in Table~\ref{tab:dm_property_performance}. Full automatic differentiation (AD) obtains the density and its gradient from second- and third-order spatial derivatives of $V_H$. The resulting graph increases memory and runtime, and its numerical stability deteriorates on the large grids required by large molecules. We therefore introduce Hybrid for faster and stable inference.

{First is the hybridization of numerical gradient and auto differentiation.} Hybrid uses a seven-point central finite difference to obtain the learned local density and first-order AD of that density to obtain $\nabla\rho$. {A scan of ten candidate steps on 100 QM9 test molecules selected 0.03 Bohr by the lowest mean wNMAE relative to full AD (Figure~\ref{fig:hybrid-step-scan}, \ref{app:hybrid-step}).} With Cartesian unit vectors $\mathbf e_\alpha$ and $h=0.03$ Bohr,

\begin{align}
	\nabla_h^2\Phi_\theta(\mathbf r)
	&=\frac{1}{h^2}\sum_{\alpha\in\{x,y,z\}}
	\left[\Phi_\theta(\mathbf r+h\mathbf e_\alpha)-2\Phi_\theta(\mathbf r)
	+\Phi_\theta(\mathbf r-h\mathbf e_\alpha)\right], \\
	\rho_{\mathrm{Hybrid}}(\mathbf r)
	&=-\frac{1}{4\pi}\nabla_h^2\Phi_\theta(\mathbf r)
	+\sum_{i,k} q_{ik}\,G_{ik}(\mathbf r),
\end{align}

where $G_{ik}$ denotes the Gaussian density. The indices $i$ and $k$ identify atoms and their Gaussian shells, and $q_{ik}$ gives each shell charge. These shells include the fixed atomic prior and the environment-dependent shells predicted by the network. The corresponding Hartree potential satisfies $\rho_G=-\nabla^2V_{\mathrm G}/(4\pi)$, so closed-form expressions provide its density and gradient. Hybrid applies the finite-difference stencil only to $\Phi_\theta$, then uses first-order AD to differentiate the resulting density. This route removes the second- and third-order reverse-mode graphs used by full AD.

{Secondly, we make the hybridization of analytical integral, high level and low level numerical integral.The Query Point Model, although theoretically scalable with a local truncation, exhibits a computational burden that grows sharply with the number of points. By separating the different components of the potential and performing calculations according to the characteristics of each component, we significantly reduce the number of points required for the query point part, thereby enhancing the computational speed of the Hybrid method, i.e., the hybridization of different computational methods.}

{Specifically, the effective potential provided by the model can be decomposed into global, local Hartree potential, and exchange-correlation potential.}

\begin{equation}
	v_{\text{eff}} = v_H^{\text{global}} + v_H^{\text{local}} + v_{xc}[\hat \rho_{\text{global}}+\hat \rho_{\text{local}}]
\end{equation}

{Separating the analytical Gaussian Hartree contribution leaves a smoother local potential, whose matrix elements are evaluated on a level-0 atom-centred quadrature grid. For the XC contribution, we first evaluate the full predicted density on the same coarse grid. We then estimate the correction from level 0 to level 3 using the inexpensive Gaussian density. Specifically, we add the difference between the Gaussian-density XC matrices evaluated at levels 3 and 0 to the full-density level-0 matrix. This approximation assumes that the Gaussian component captures the dominant grid dependence of the full-density XC contribution, particularly near the nuclei. The strategy is analogous to correcting a small-basis CCSD(T) energy using the difference between large- and small-basis MP2 energies. Here, grid refinement replaces basis-set enlargement, and the Gaussian density provides the inexpensive reference for estimating the correction. This avoids evaluating the neural-network density on the finer grid. The resulting effective potential matrix is given by Eq.~\ref{eq:hybrid_effective_potential}.}

\begin{equation}
{
\begin{aligned}
\bigl[\widehat{\mathbf V}_{\mathrm{eff}}^{\mathrm{hyb}}\bigr]_{\mu\nu}
={}&
\sum_{i,k}q_{ik}(\mu\nu\mid ik)
+\bigl[\widehat{\mathbf J}_{\mathrm{local}}^{(0)}\bigr]_{\mu\nu}
+\bigl[\mathbf V_{\mathrm{xc}}^{(0)}[\widehat\rho_{\mathrm{total}}]\bigr]_{\mu\nu}
\\
&+\bigl[\mathbf V_{\mathrm{xc}}^{(3)}[\widehat\rho_{\mathrm G}]
-\mathbf V_{\mathrm{xc}}^{(0)}[\widehat\rho_{\mathrm G}]\bigr]_{\mu\nu},
\\[3pt]
(\mu\nu\mid ik)
={}&\iint
\frac{\chi_\mu(\mathbf r)\chi_\nu(\mathbf r)G_{ik}(\mathbf r')}
{|\mathbf r-\mathbf r'|}
\,\mathrm d\mathbf r\,\mathrm d\mathbf r'.
\end{aligned}
}
\label{eq:hybrid_effective_potential}
\end{equation}

{Here, $(\mu\nu\mid ik)$ is the three-centre two-electron Coulomb integral over orbital-basis functions $\chi_\mu$, $\chi_\nu$ and the unit-integral density Gaussian $G_{ik}$. The coefficient $q_{ik}$ is the shell charge, and the sum includes both fixed and learnable Gaussian shells.}

{In this section, the energy and each energy component are recalculated using the first diagonalization density matrix. Orbital properties such as HOMO, LUMO, and Gap are derived from the first diagonalization. The dipole moment can be derived using the latent charge obtained from density inference, which can be proven by the divergence theorem. The energy-evaluation formulas and dipole derivation are given in \ref{app:energy} and \ref{app:dipole}, respectively.}

We compared occupied spaces using the spin-summed AO density matrices $D_{\mathrm{model}}$ and $D_{\mathrm{ref}}$ and the AO overlap matrix $S$,

\begin{equation}
	O_{\mathrm{occ}}
	=\frac{\operatorname{Tr}\!\left(D_{\mathrm{model}}S D_{\mathrm{ref}}S\right)}{2N_{\mathrm e}}
	=\frac{1}{N_{\mathrm{occ}}}
	\left\|C_{\mathrm{occ}}^{\mathrm{model}\,\mathsf T}
	S C_{\mathrm{occ}}^{\mathrm{ref}}\right\|_{\mathrm F}^{2}.
\end{equation}

Here $C_{\mathrm{occ}}$ contains the occupied molecular-orbital coefficients and $N_{\mathrm{occ}}=N_{\mathrm e}/2$. This metric equals one for identical occupied subspaces and is invariant to rotations among occupied orbitals.

{To measure how close the predicted density is to a Kohn--Sham fixed point, we define an index $\delta_\text{FP}$, which calculates the degree of agreement between the derived density of the density matrix obtained from the first diagonalization and the original neural field density.}

\begin{equation}
	{
	\delta_{\rm FP}
   =\frac{1}{N_e}\int|\hat\rho(\mathbf r)-\rho[D^{(0)}](\mathbf r)|\,d\mathbf r
	}
\end{equation}

\begin{table*}[!htbp]
{
\centering
\caption{{
Performance of the AD and Hybrid methods across different orbital basis sets.
Results are reported for all 13,389 molecules in the QM9 test set;
Panel (a) reports mean absolute errors (MAEs), and panel (b) reports coefficients
of determination ($R^2$).
The $R^2$ values of $\lvert 1-O_{\mathrm{occ}}\rvert$ and
$\delta_{\mathrm{FP}}$ are not reported because both error quantities have
constant zero reference values.
$\delta_{\mathrm{FP}}$ is not reported for Hybrid because its total model
density is evaluated on the level-0 coarse grid rather than on the complete level-3 grid.
The dipole moment of the Hybrid is the same as that of the AD converter, and both use the same algorithm, so it will not be reported again.
}
}
\label{tab:dm_property_performance}

\renewcommand{\arraystretch}{1.12}
\setlength{\tabcolsep}{2pt}

% ============================================================
% Panel a: MAE
% ============================================================
\begin{minipage}[t]{0.45\textwidth}
\vspace{0pt}
\centering
{\small\bfseries (a) Mean absolute error (MAE)}
\par\smallskip

\scriptsize
\resizebox{\linewidth}{!}{%
\begin{tabular}{
    @{}
    >{\raggedright\arraybackslash}p{2.6cm}
    *{6}{>{\centering\arraybackslash}p{0.9cm}}
    @{}
}
\toprule
\multicolumn{1}{c}{Property}
& \multicolumn{2}{c}{def2-SVP}
& \multicolumn{2}{c}{def2-TZVP}
& \multicolumn{2}{c}{def2-QZVP} \\
\cmidrule(lr){2-3}
\cmidrule(lr){4-5}
\cmidrule(lr){6-7}
& AD & Hybrid
& AD & Hybrid
& AD & Hybrid \\
\midrule
$E_{\mathrm{tot}}$ (meV atom$^{-1}$)
    & 10.63 & 12.65 & 10.64 & 13.38 & 10.89 & 13.56 \\
$E_{\mathrm{kin}}$ (meV atom$^{-1}$)
    & 538.38 & 597.04 & 914.47 & 987.39 & 884.99 & 966.70 \\
$E_{\mathrm{ne}}$ (meV atom$^{-1}$)
    & 1484.01 & 1633.37 & 1887.22 & 2284.92 & 1885.97 & 2288.76 \\
$E_J$ (meV atom$^{-1}$)
    & 1033.08 & 1119.25 & 1135.16 & 1449.51 & 1155.57 & 1470.63 \\
$E_{\mathrm{xc}}$ (meV atom$^{-1}$)
    & 71.41 & 82.52 & 98.82 & 121.54 & 99.89 & 122.70 \\
HOMO (meV)
    & 272.39 & 315.42 & 230.48 & 264.52 & 219.89 & 250.00 \\
LUMO (meV)
    & 151.93 & 149.26 & 144.09 & 158.62 & 149.03 & 175.64 \\
Gap (meV)
    & 210.09 & 304.65 & 172.70 & 258.06 & 176.62 & 258.10 \\
$|\mu|$ (D)
    & 0.805 & -- & 0.907 & -- & 0.931 & -- \\
$\lvert 1-O_{\mathrm{occ}}\rvert$ ($10^{-5}$)
    & 6.762 & 8.549 & 6.708 & 8.963 & 6.863 & 9.091 \\
$\delta_{\mathrm{FP}}$ (\%)
    & 1.801 & -- & 2.991 & -- & 3.007 & -- \\
\bottomrule
\end{tabular}%
}
\end{minipage}%
\hspace{0.05\textwidth}%
% ============================================================
% Panel b: R-squared
% ============================================================
\begin{minipage}[t]{0.45\textwidth}
\vspace{0pt}
\centering
{\small\bfseries (b) Coefficient of determination ($R^2$)}
\par\smallskip

\scriptsize
\resizebox{\linewidth}{!}{%
\begin{tabular}{
    @{}
    >{\raggedright\arraybackslash}p{2.6cm}
    *{6}{>{\centering\arraybackslash}p{0.9cm}}
    @{}
}
\toprule
\multicolumn{1}{c}{Property}
& \multicolumn{2}{c}{def2-SVP}
& \multicolumn{2}{c}{def2-TZVP}
& \multicolumn{2}{c}{def2-QZVP} \\
\cmidrule(lr){2-3}
\cmidrule(lr){4-5}
\cmidrule(lr){6-7}
& AD & Hybrid
& AD & Hybrid
& AD & Hybrid \\
\midrule
$E_{\mathrm{tot}}$
    & 1.000 & 1.000 & 1.000 & 1.000 & 1.000 & 1.000 \\
$E_{\mathrm{kin}}$
    & 1.000 & 1.000 & 1.000 & 1.000 & 1.000 & 1.000 \\
$E_{\mathrm{ne}}$
    & 1.000 & 1.000 & 1.000 & 1.000 & 1.000 & 1.000 \\
$E_J$
    & 1.000 & 1.000 & 1.000 & 1.000 & 1.000 & 1.000 \\
$E_{\mathrm{xc}}$
    & 1.000 & 1.000 & 1.000 & 1.000 & 1.000 & 1.000 \\
HOMO
    & 0.646 & 0.537 & 0.726 & 0.660 & 0.734 & 0.682 \\
LUMO
    & 0.963 & 0.962 & 0.957 & 0.952 & 0.943 & 0.935 \\
Gap
    & 0.952 & 0.904 & 0.959 & 0.914 & 0.950 & 0.899 \\
$|\mu|$
    & 0.613 & -- & 0.553 & -- & 0.531 & -- \\
$\lvert 1-O_{\mathrm{occ}}\rvert$
    & -- & -- & -- & -- & -- & -- \\
$\delta_{\mathrm{FP}}$
    & -- & -- & -- & -- & -- & -- \\
\bottomrule
\end{tabular}%
}
\end{minipage}
}
\end{table*}

\paragraph{QM9}

\par\begingroup
On QM9, the continuous fields learned from PBE/def2-SVP densities recover both energetic and orbital information through a single Kohn--Sham diagonalization. In the training basis, AD yields near-unit $R^2$ for the total energy and its components (Fig.~\ref{fig:qm9_ad_properties}\textbf{a}--\textbf{e}). Agreement varies more across the frontier orbital properties: the HOMO has a lower $R^2$ than the LUMO and HOMO--LUMO gap, both of which exceed 0.95 (Fig.~\ref{fig:qm9_ad_properties}\textbf{f}--\textbf{h}). The dipole-moment magnitude, obtained directly from the model-derived atomic charges, shows weaker agreement (Fig.~\ref{fig:qm9_ad_properties}\textbf{i}). These results establish the accuracy of the learned field and its conversion to electronic structure within the training basis.

The same QM9 model transfers directly to def2-TZVP and def2-QZVP without retraining. We project its predicted fields into each target AO basis and compare the resulting properties with converged PBE references in that basis (Table~\ref{tab:dm_property_performance}). The AD total-energy MAE stays within 10.63--10.89~meV atom$^{-1}$ across all three bases, while the occupied-subspace deficit remains near $7\times10^{-5}$. HOMO, LUMO, and gap MAEs are lower in both unseen bases than in def2-SVP. Individual energy-component errors and $\delta_{\mathrm{FP}}$ increase relative to the training basis, indicating greater sensitivity to the AO representation. The preserved total-energy accuracy and occupied-subspace agreement demonstrate that the learned field supports electronic-structure reconstruction beyond the basis used to generate its training targets.

Hybrid retains this cross-basis capability through def2-QZVP, with total-energy errors close to those of AD and small occupied-subspace deficits across all three bases. Its numerical approximations have a more pronounced effect on frontier orbital spacings: the gap MAE is approximately 45--49\% higher than that of AD. This consistent pattern identifies a property-dependent trade-off in the Hybrid conversion, which preserves overall energetic accuracy across bases while introducing larger HOMO--LUMO gap errors.
\par\endgroup

\begin{table*}[!htbp]
{
\centering
\caption{{
Performance of the AD and Hybrid methods across different orbital basis sets
on the VQM24 test set.
Results are reported for all 25,825 VQM24 test molecules.
Panel (a) reports mean absolute errors (MAEs), and panel (b) reports
coefficients of determination ($R^2$).
Dashes denote quantities that are not reported.
}}
\label{tab:vqm24_dm_property_performance}

\renewcommand{\arraystretch}{1.12}
\setlength{\tabcolsep}{2pt}

% ============================================================
% Panel a: MAE
% ============================================================
\begin{minipage}[t]{0.45\textwidth}
\vspace{0pt}
\centering
{\small\bfseries (a) Mean absolute error (MAE)}
\par\smallskip

\scriptsize
\resizebox{\linewidth}{!}{%
\begin{tabular}{
@{}
>{\raggedright\arraybackslash}p{2.6cm}
*{6}{>{\centering\arraybackslash}p{0.9cm}}
@{}
}
\toprule
\multicolumn{1}{c}{Property}
& \multicolumn{2}{c}{def2-SVP}
& \multicolumn{2}{c}{def2-TZVP}
& \multicolumn{2}{c}{def2-QZVP} \\
\cmidrule(lr){2-3}
\cmidrule(lr){4-5}
\cmidrule(lr){6-7}
& AD & Hybrid
& AD & Hybrid
& AD & Hybrid \\
\midrule
$E_{\mathrm{tot}}$ (meV atom$^{-1}$)
& 99.36 & 99.79 & 94.63 & 101.00 & 97.78 & 104.69 \\
$E_{\mathrm{kin}}$ (meV atom$^{-1}$)
& 6115.29 & 5456.62 & 6032.76 & 5422.61 & 5181.02 & 4681.37 \\
$E_{\mathrm{ne}}$ (eV atom$^{-1}$)
& 15.59 & 13.65 & 14.90 & 12.81 & 14.02 & 12.14 \\
$E_J$ (eV atom$^{-1}$)
& 10.05 & 8.68 & 9.43 & 7.87 & 9.43 & 7.94 \\
$E_{\mathrm{xc}}$ (meV atom$^{-1}$)
& 668.93 & 584.37 & 659.48 & 561.98 & 646.45 & 551.94 \\
HOMO (meV)
& 312.50 & 291.97 & 311.80 & 287.27 & 321.92 & 292.31 \\
LUMO (meV)
& 440.60 & 542.99 & 383.74 & 490.06 & 398.53 & 505.33 \\
Gap (meV)
& 364.03 & 524.49 & 297.34 & 446.57 & 288.64 & 432.57 \\
$|\mu|$ (D)
& 0.593 & -- & 0.577 & -- & 0.578 & -- \\
$\lvert 1-O_{\mathrm{occ}}\rvert$ ($10^{-5}$)
& 46.708 & 48.260 & 45.396 & 48.516 & 46.570 & 50.045 \\
$\delta_{\mathrm{FP}}$ (\%)
& 3.178 & -- & 3.589 & -- & 3.649 & -- \\
\bottomrule
\end{tabular}%
}
\end{minipage}%
\hspace{0.05\textwidth}%
% ============================================================
% Panel b: R-squared
% ============================================================
\begin{minipage}[t]{0.45\textwidth}
\vspace{0pt}
\centering
{\small\bfseries (b) Coefficient of determination ($R^2$)}
\par\smallskip

\scriptsize
\resizebox{\linewidth}{!}{%
\begin{tabular}{
@{}
>{\raggedright\arraybackslash}p{2.6cm}
*{6}{>{\centering\arraybackslash}p{0.9cm}}
@{}
}
\toprule
\multicolumn{1}{c}{Property}
& \multicolumn{2}{c}{def2-SVP}
& \multicolumn{2}{c}{def2-TZVP}
& \multicolumn{2}{c}{def2-QZVP} \\
\cmidrule(lr){2-3}
\cmidrule(lr){4-5}
\cmidrule(lr){6-7}
& AD & Hybrid
& AD & Hybrid
& AD & Hybrid \\
\midrule
$E_{\mathrm{tot}}$
& 1.000 & 1.000 & 1.000 & 1.000 & 1.000 & 1.000 \\
$E_{\mathrm{kin}}$
& 1.000 & 1.000 & 1.000 & 1.000 & 1.000 & 1.000 \\
$E_{\mathrm{ne}}$
& 1.000 & 1.000 & 1.000 & 1.000 & 1.000 & 1.000 \\
$E_J$
& 1.000 & 1.000 & 1.000 & 1.000 & 1.000 & 1.000 \\
$E_{\mathrm{xc}}$
& 1.000 & 1.000 & 1.000 & 1.000 & 1.000 & 1.000 \\
HOMO
& 0.535 & 0.591 & 0.526 & 0.599 & 0.490 & 0.579 \\
LUMO
& 0.731 & 0.623 & 0.747 & 0.627 & 0.717 & 0.587 \\
Gap
& 0.829 & 0.681 & 0.869 & 0.736 & 0.871 & 0.740 \\
$|\mu|$
& 0.642 & -- & 0.663 & -- & 0.660 & -- \\
$\lvert 1-O_{\mathrm{occ}}\rvert$
& -- & -- & -- & -- & -- & -- \\
$\delta_{\mathrm{FP}}$
& -- & -- & -- & -- & -- & -- \\
\bottomrule
\end{tabular}%
}
\end{minipage}
}
\end{table*}

\paragraph{VQM24.}
\par\begingroup
The independently trained VQM24 model tests the same field-to-matrix construction across a broader range of elemental environments, including Si, P, S, Cl, and Br. With training again restricted to PBE/def2-SVP densities, AD reproduces the overall energy trends in the training basis despite larger absolute errors than on QM9 (Table~\ref{tab:vqm24_dm_property_performance}). The total-energy MAE is 99.36~meV atom$^{-1}$, and the total energy and its components retain near-unit $R^2$. Frontier orbital predictions show greater variation in accuracy, with the HOMO less strongly correlated with its reference than the LUMO and gap. The occupied-subspace deficit and $\delta_{\mathrm{FP}}$ are also larger than on QM9, reflecting the greater difficulty of reconstructing electronic structure across this chemical space.

Cross-basis transfer remains effective in this broader setting. Applying the unchanged VQM24 model in def2-TZVP and def2-QZVP keeps the AD total-energy MAE below 100~meV atom$^{-1}$ and preserves occupied-subspace agreement. Both unseen bases yield lower LUMO and gap MAEs than def2-SVP, with the gap $R^2$ reaching 0.871 in def2-QZVP, while $\delta_{\mathrm{FP}}$ rises modestly. Together with the QM9 results, this demonstrates that fields learned from def2-SVP density targets support one-shot reconstruction in both triple- and quadruple-zeta AO bases.

Hybrid exhibits a distinct balance of errors on VQM24. Across all three orbital bases, it lowers the energy-component and HOMO MAEs relative to AD while leaving the total-energy MAE close to the AD value. The LUMO and gap are more sensitive to the approximation, with gap MAEs increasing by approximately 44--50\%. This pattern persists through def2-QZVP, showing that Hybrid retains cross-basis transfer with a larger accuracy cost for the unoccupied frontier level and its separation from the HOMO.
\par\endgroup

\noindent\paragraph{{Representative spectra.}}

\par
\begin{figure}[h]
	\centering
	\includegraphics[width=\linewidth]{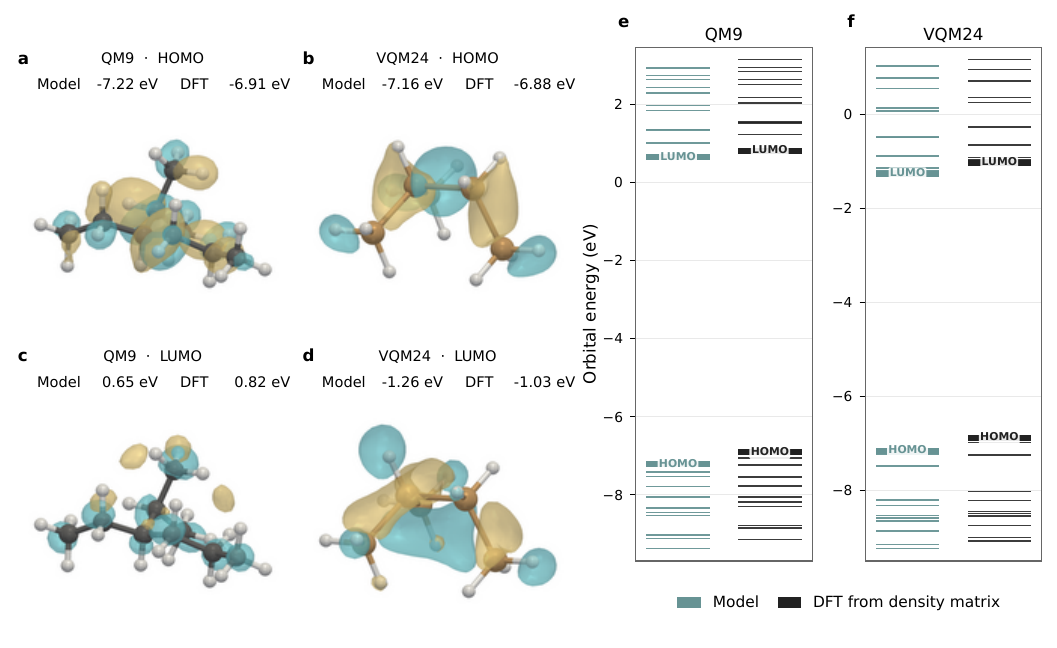}
	\caption{\textbf{Representative frontier orbitals and energy spectra for the largest test molecules.} {Panels \textbf{a}--\textbf{d} show one-shot AD HOMO and LUMO isosurfaces for the largest molecule by atom count in each test set. The molecules are QM9 \texttt{gdb\_118466} ($\mathrm{C}_9\mathrm{H}_{20}$, 29 atoms) and VQM24 \texttt{vqm24\_256936} ($\mathrm{Si}_5\mathrm{H}_{12}$, 17 atoms), evaluated with their respective dataset-trained models.} Labels report the corresponding one-shot and reference DFT orbital energies. Panels \textbf{e} and \textbf{f} compare 22 index-matched levels from HOMO$-10$ through LUMO$+10$. Teal denotes DeepHartree, and black denotes the converged density-fitted PBE/def2-SVP reference.}
	\label{fig:largest_test_orbitals}
\end{figure}

\begingroup
The frontier-orbital surfaces show the spatial localization and nodal structure of the one-shot states (Fig.~\ref{fig:largest_test_orbitals}\textbf{a}--\textbf{d}). In both representative molecules, QM9 \texttt{gdb\_118466} and VQM24 \texttt{vqm24\_256936}, the model HOMO and LUMO energies lie below the DFT values. The extended spectra show the same downward displacement across all 22 frontier-adjacent, index-matched levels (Fig.~\ref{fig:largest_test_orbitals}\textbf{e},\textbf{f}). Subtracting the molecule-specific mean shift reduces the RMS error from 0.241 to 0.045~eV for QM9 and from 0.191 to 0.054~eV for VQM24.
\par\endgroup

\par\begingroup
For these two molecules, a nearly uniform downward energy shift dominates the spectral error, while the relative level pattern is preserved. The common HOMO and LUMO shift partially cancels in their separation, yielding gap errors of 0.137~eV for QM9 and 0.039~eV for VQM24.
\par\endgroup

\subsection{Learned initialization reduces SCF iterations}

{We constructed DeepHartree density matrices using the Hybrid multilevel scheme in Eq.~\ref{eq:hybrid_effective_potential}, which combines analytical Gaussian Coulomb integrals with level-0 quadrature and a level-3 Gaussian-density XC correction. The subsequent GPU4PySCF RKS calculations used level-3 atom-centred grids \citep{li2025gpu4pyscf}.} We evaluated these density matrices as initial guesses for all 13,389 QM9 test molecules {under three different bases(def2-SVP,def2-TZVP,def2-QZVP)}. We compared each converged iteration count with the corresponding minimal atomic-orbital (MINAO) reference stored in the dataset metadata. All model-initialized calculations converged. For molecule $i$, we define the relative iteration count (RIC) as
\begin{equation}
    \mathrm{RIC}_i=\frac{N_i^{\mathrm{hybrid}}}{N_i^{\mathrm{ref}}}.
    \label{eq:ric}
\end{equation}

\begin{figure}[!t]
	\centering
	\includegraphics[width=\linewidth]{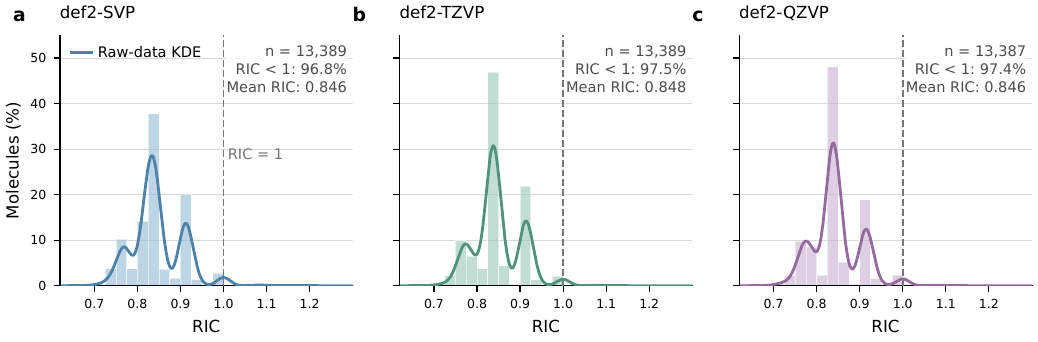}
	\caption{{\textbf{DeepHartree initialization reduces SCF iterations across three basis sets.} Panels \textbf{a}--\textbf{c} show def2-SVP, def2-TZVP, and def2-QZVP, respectively, for the QM9 test molecules. Bars show the percentage of converged, directly comparable molecules per 0.025-wide RIC bin. Solid curves are Gaussian kernel estimates from individual RIC values (bandwidth 0.015 RIC units), scaled to the same percentage-per-bin axis. The dashed line marks RIC = 1. The two def2-QZVP references repaired with Newton/SOSCF are excluded, leaving $n=13{,}389$, $13{,}389$, and $13{,}387$ molecules in panels \textbf{a}--\textbf{c}, respectively.}}
	\label{fig:scf_iteration_reduction}
\end{figure}

\par\begingroup
Mean SCF iterations decreased from 12.215 to 10.325 with def2-SVP, from 12.511 to 10.596 with def2-TZVP, and from 12.652 to 10.684 with def2-QZVP (Fig.~\ref{fig:scf_iteration_reduction}). These changes correspond to reductions of 15.5\%, 15.3\%, and 15.6\% in the respective mean iteration counts. The mean molecule-level RIC values were 0.846, 0.848, and 0.846. Iteration counts decreased for 12,965 of 13,389 molecules (96.8\%) with def2-SVP, 13,055 of 13,389 (97.5\%) with def2-TZVP, and 13,045 of 13,387 (97.4\%) with def2-QZVP. These similar distributions show that the iteration reduction persists across all three AO bases.
\par\endgroup

\par
\begin{figure}[!htbp]
	\centering
	\includegraphics[width=0.5\linewidth]{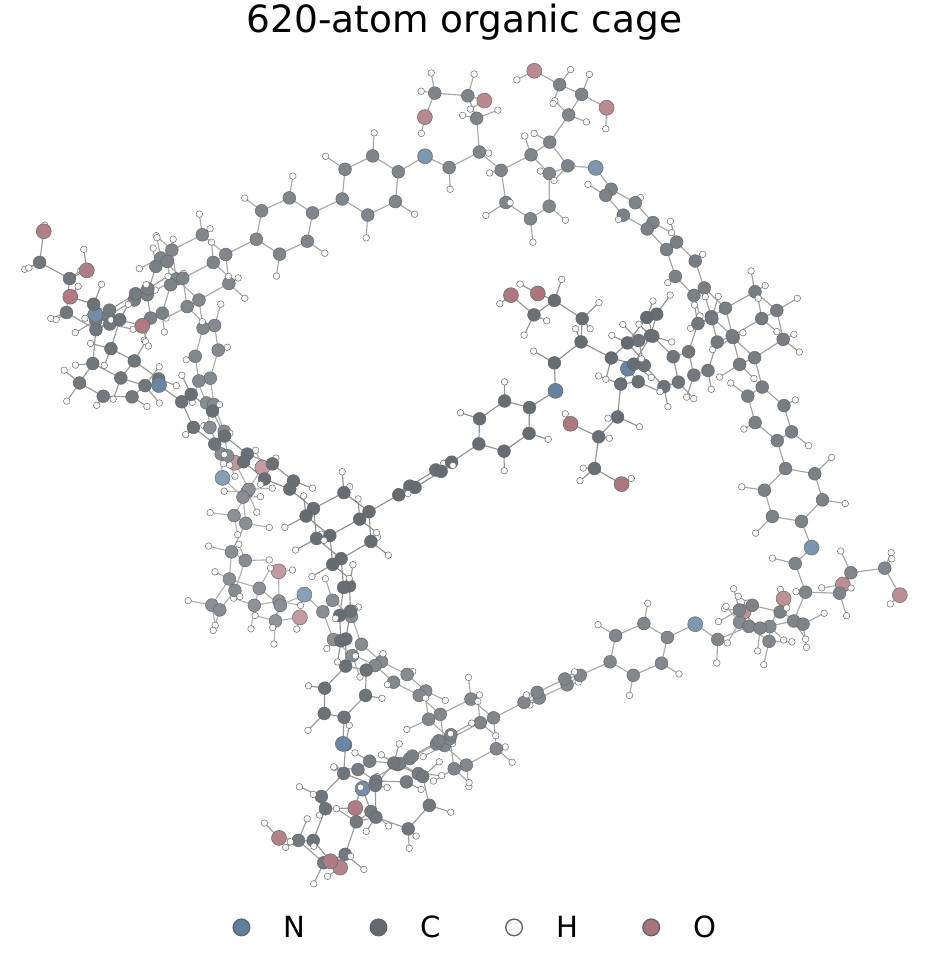}
	\caption{{\textbf{Organic cage used for the large-system SCF wall-clock benchmark.} The 620-atom structure contains 280 carbon, 304 hydrogen, 12 nitrogen, and 24 oxygen atoms. Gray, white, blue, and rose spheres denote these elements, respectively. The coordinates are from the VAE-generated Cage-VAE structure; RDKit inferred the displayed connections from the XYZ coordinates.}}
	\label{fig:large_system_cage}
\end{figure}

\begingroup
The model scales favorably with system size. We therefore tested whether using it for initialization yields an end-to-end wall-clock speedup for SCF on a larger system. We benchmarked a 620-atom organic cage generated by Cage-VAE \citep{zhou2023cagevae} (Fig.~\ref{fig:large_system_cage}). On a single NVIDIA RTX 5090, we compared model-initialized and MINAO-initialized PBE/def2-SVP calculations using GPU4PySCF and measured each stage of model-based guess construction.
\par\endgroup

\par\begingroup
Preparing the model-based initial guess took 174.63~s. Coarse-grid construction (531,456 points) took 8.10~s, and coarse-grid density inference took 37.33~s. Fine-grid construction (7,439,616 points) took 63.43~s, the largest preparation component; fine-grid density inference took 18.95~s. Matrix construction took 30.46~s, including 15.92~s for analytic Coulomb-matrix evaluation. Diagonalization and setup took 6.94 and 9.42~s, respectively. With this guess, SCF converged in 16 iterations over 629.23~s, giving a total of 803.86~s. MINAO-initialized SCF converged in 20 iterations over 922.94~s, with a total of 922.95~s. Model-based initialization therefore saved 119.09~s overall, reducing wall-clock time by 12.9\% (1.15-fold speedup).
\par\endgroup

\FloatBarrier

\subsection{Out of distribution test}
\label{sec:oe62_ood}

\begin{figure}[!htbp]
	\centering
	\includegraphics[width=\linewidth]{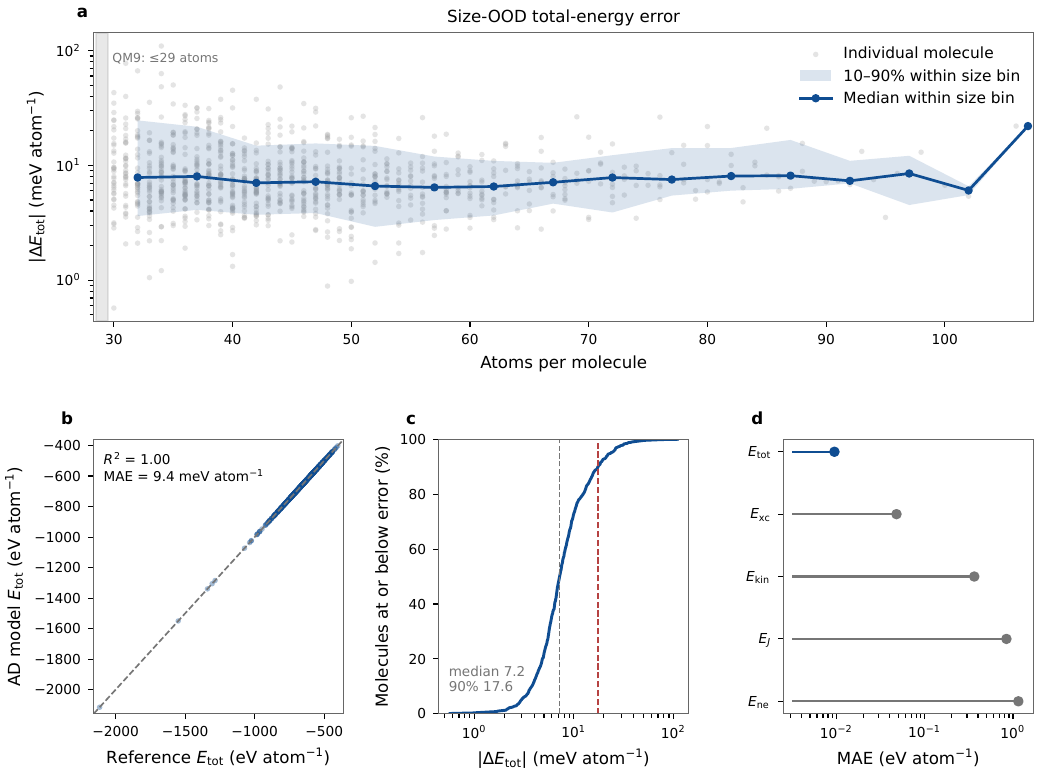}
	\caption{\textbf{The QM9-trained model preserves total energy on larger OE62 molecules.} Panel \textbf{a} reports the absolute total-energy error against atom count; points denote molecules, the line denotes the within-bin median, and the band spans the 10th--90th percentiles. Panel \textbf{b} compares model and reference total energies per atom. Panel \textbf{c} gives the cumulative absolute-error distribution. Panel \textbf{d} reports MAEs for the total energy and individual energy components. The 1,000 H/C/N/O/F molecules contain 30--106 atoms and lie beyond the QM9 size range.}
	\label{fig:oe62_ood_generalization}
\end{figure}

We evaluated size out-of-distribution (OOD) generalization on OE62 while keeping the QM9 element set \citep{stuke2020atomic}. We retained H/C/N/O/F molecules with more than 29 atoms, the largest QM9 size, and sampled 1,000 molecules in proportion to their exact-atom-count strata. The resulting set spans 30--106 atoms and 71 represented atom counts.The evaluation used the QM9-trained PaiNN checkpoint without OE62 fine-tuning. 

The QM9-trained model reaches a total-energy MAE of {9.437}~meV atom$^{-1}$ on OE62, with a median of {7.185}~meV atom$^{-1}$ and a 90th percentile of {17.552}~meV atom$^{-1}$ (Fig.~\ref{fig:oe62_ood_generalization}\textbf{a}--\textbf{c}). The per-atom total energy retains {$R^2=0.99999999$.} Across the complete 30--106-atom range, the size-binned median remains near 10~meV atom$^{-1}$ and shows no systematic growth with molecular size.

The component MAEs are {0.048, 0.364, 0.843,} and {1.151} ~eV atom$^{-1}$ for exchange--correlation, kinetic, Coulomb, and electron--nuclear energies, respectively (Fig.~\ref{fig:oe62_ood_generalization}\textbf{d}). {These values are 18--33\% below the corresponding QM9 AD/def2-SVP errors in Table~\ref{tab:dm_property_performance}, demonstrating that component-wise accuracy is retained under size extrapolation.} Cancellation among the larger component errors reduces the total-energy MAE to {9.437}~meV atom$^{-1}$. The OE62 test demonstrates size-OOD transfer of the one-shot energy decomposition, with total-energy accuracy benefiting strongly from error cancellation.

\noindent\subsection{{Scaling of one-shot density-matrix construction}}
\label{sec:scaling}

\par\begingroup
To quantify the computational cost of extending one-shot prediction to larger systems, we benchmarked all-trans linear alkanes (77--1001 atoms) and poly(ethylene glycol) (PEG) chains (80--1004 atoms). We compared AD and Hybrid density-matrix construction with PBE/def2-SVP SCF calculations in CPU PySCF and GPU4PySCF, with and without density fitting. CPU calculations used 32 Intel Xeon Platinum 8473C threads; GPU calculations used one NVIDIA RTX 4090. Timings exclude grid construction, and scaling exponents were obtained by fitting mean times to $t=A N_{\mathrm{atom}}^n$ in log--log space. The AD implementation was optimized for large systems. For molecules containing more than 400 Gaussian shells, the Gaussian Hartree potential was decomposed into long- and short-range components. The fast multipole method evaluated the long-range point-charge contribution, while distance-based screening localized the short-range correction, Gaussian density, and density gradient. Further implementation details are provided in \ref{app:screening}.
\par\endgroup

\begin{figure}[!htbp]
    \centering
    \includegraphics[width=\linewidth]{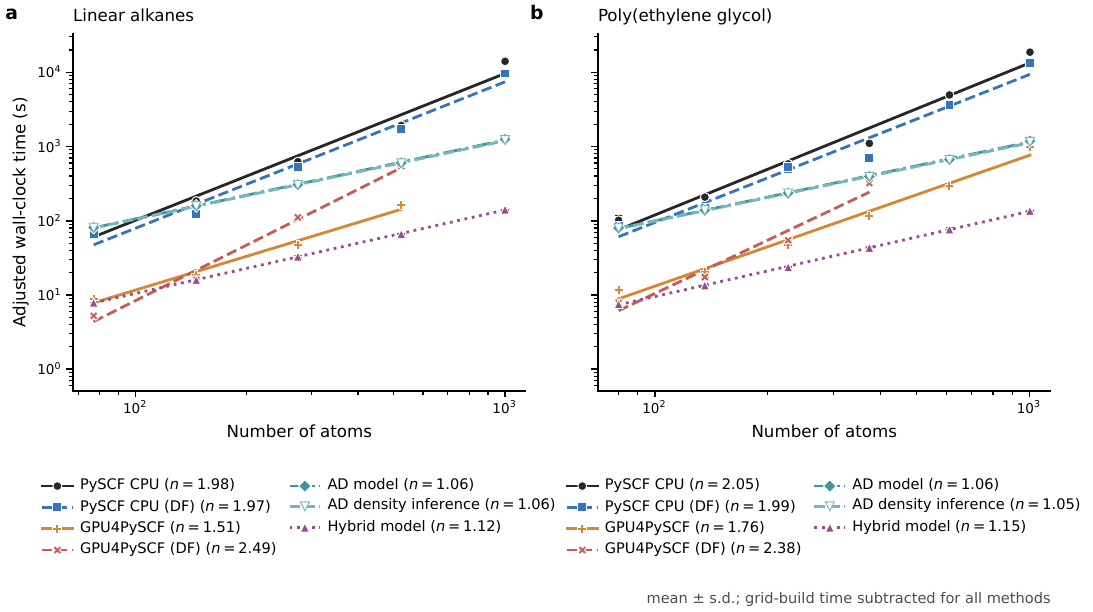}
    \caption{\noindent{\textbf{DeepHartree density-matrix construction scales nearly linearly with molecular size.} Panels \textbf{a} and \textbf{b} show all-trans linear alkanes and PEG chains, respectively. Timings exclude grid construction. Points show mean times. Lines are log--log ordinary-least-squares fits, with exponents $n$ reported in the legends. The separate AD density-inference series reports molecular encoding and field evaluation.}}
    \label{fig:scaling_wall_clock}
    \label{fig:img/scaling_wall_clock_loglog.pdf}
\end{figure}

\par\begingroup
DeepHartree density-matrix construction exhibited near-linear scaling in both molecular families (Fig.~\ref{fig:img/scaling_wall_clock_loglog.pdf}). The AD exponents were 1.062 for alkanes and 1.060 for PEG, and the Hybrid exponents were 1.123 and 1.146, respectively. CPU PySCF exponents ranged from 1.971 to 2.047 across both families and density-fitting settings, while GPU4PySCF exponents ranged from 1.512 to 2.486. The dominant computational cost was model density inference at quadrature query points. For the 1001-atom alkane, density inference accounted for 97.5\% of AD runtime and 67.2\% of Hybrid runtime. Matrix diagonalization was not a bottleneck within the tested size range.
\par\endgroup

\par\begingroup
Hybrid combined near-linear scaling with the shortest model runtime. For the 1001-atom alkane, it constructed the density matrix in 141.8~s, an 8.84-fold speedup over AD. The corresponding speedups over CPU PySCF were 99.1-fold without density fitting and 68.0-fold with density fitting. At 524 atoms, Hybrid was 2.43-fold faster than GPU4PySCF and 8.27-fold faster than its density-fitted variant. For the 1004-atom PEG chain, Hybrid required 135.6~s, giving an 8.75-fold speedup over AD and a 7.34-fold speedup over GPU4PySCF. These results establish efficient one-shot density-matrix construction for systems containing up to 1004 atoms, with Hybrid delivering the largest computational gains.
\par\endgroup

% !TEX root = ../main.tex
\section{Methods}

The DeepHartree framework implements a Poisson-coupled neural field: an E(3)-equivariant graph neural network predicts the Hartree potential $V_H(\mathbf{r})$ as a continuous scalar field, and the electron density is obtained exactly through the Poisson equation, $\rho(\mathbf{r}) = -\nabla^2 V_H(\mathbf{r}) / 4\pi$. Below we describe graph construction, equivariant message passing, singularity removal via delta-learning, model training, and one-shot density matrix construction from the predicted fields.

\subsection{Graph construction}

To efficiently predict electron density, we model the system as a heterogeneous directed graph $\mathcal{G} = (\mathcal{V}, \mathcal{E})$. The vertex set $\mathcal{V} = \mathcal{V}_A \cup \mathcal{V}_G$ contains atomic nodes ($N_A$) and spatial mesh nodes ($N_G$). After message passing, the final atom features parameterize element-specific Gaussian coefficients and widths. The final grid features are decoded directly into the local Hartree-potential correction $\Phi_\theta(\mathbf r_p)$.

In view of the actual system $N_G \gg N_A$, in order to avoid the huge computational overhead caused by full connectivity and to follow the principle of spatial locality, we set the cutoff radius to $R_c = 5~\text{\AA}$ when constructing the edge set $\mathcal{E}_{AA} \cup \mathcal{E}_{AG}$. Specifically, $\mathcal{E}_{AA}$ is the bidirectional edge between atoms within $R_c$ to capture many-body interactions and local chemical environments; $\mathcal{E}_{AG}$ is the unidirectional edge of the grid from the atoms to the grid within $R_c$. To accurately describe these two distinct physical mapping processes, the messaging mechanism on $\mathcal{E}_{AA}$ and $\mathcal{E}_{AG}$ employs two completely separate and unshared sets of network parameters.

In this architecture, grid nodes are strictly limited to pure information receivers (i.e., $\mathcal{E}_{GA} = \emptyset$ and $\mathcal{E}_{GG} = \emptyset$). The number of local neighbors connected to each node is limited to a constant independent of the total size of the system. This sparseness-oriented design based on physical locality strictly reduces the computational complexity of graph convolution from $\mathcal{O}(N_A^2 + N_A N_G)$ to $\mathcal{O}(N_A + N_G)$ that is linear with the system scale.

\subsection{Equivariant Graph Neural Networks and Continuous Message Passing}

\begin{figure}[t]
	\centering
	\includegraphics[width=\linewidth]{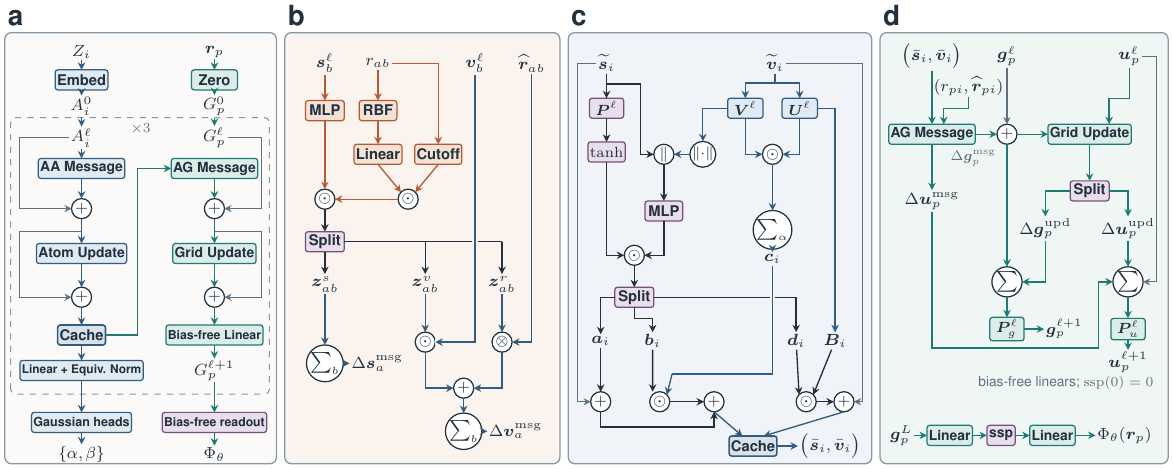}
	\caption{\textbf{The task-adapted PaiNN separates atom--atom encoding from one-way atom--grid decoding.} (\textbf{a}) The dual-stream stack alternates atom--atom messaging, atom self-updates, atom-state caching, and one-way atom--grid updates before the Gaussian and local-potential readouts. (\textbf{b}) The atom--atom and atom--grid message modules share the same scalar--vector topology but use separate weights. (\textbf{c}) The atom self-update constructs invariant gates from vector norms and inner products and caches the updated atom state. (\textbf{d}) The grid update combines cached atom messages with persistent grid states and decodes the final scalar channel through a bias-free local-potential head. Here, $A_i^\ell=(\mathbf s_i^\ell,\mathbf v_i^\ell)$ and $G_p^\ell=(\mathbf g_p^\ell,\mathbf u_p^\ell)$ denote the paired scalar--vector atom and grid states.}
	\label{fig:painn_architecture}
\end{figure}

Rigid translations, rotations, and reflections should not alter the predicted physics of a molecule: scalar-field values follow the transformed coordinates, while directional features transform with the molecular geometry. Enforcing E(3) equivariance makes this symmetry exact at every layer, avoids relearning equivalent molecular orientations, and retains directional information without introducing orientation-dependent coordinates. The predicted Hartree potential is therefore independent of the global reference frame, while internal vector channels transform covariantly.

We use a task-adapted PaiNN backbone \citep{schutt2021equivariant}. It retains PaiNN's coupled scalar--vector representation but modifies its message-passing skeleton for continuous atom-to-grid field prediction. In particular, each layer combines separate AA and AG modules, caches atom states before channel projection, propagates persistent grid states, and normalizes the atom stream equivariantly. Element-specific Gaussian heads and a bias-free grid head replace the original atom-wise property readout.

Each layer carries invariant scalar features $\mathbf s_i^\ell\in\mathbb R^{F_\ell}$ and equivariant Cartesian vector features $\mathbf v_i^\ell\in\mathbb R^{F_\ell\times3}$ on atom $i$. Grid point $p$ carries the analogous pair $(\mathbf g_p^\ell,\mathbf u_p^\ell)$, as summarized in Fig.~\ref{fig:painn_architecture}. An element embedding initializes the atomic scalars, while all atomic vectors and grid features start at zero. We use three layers with channel widths $64\rightarrow128\rightarrow128\rightarrow64$. Relative displacements remove the dependence on absolute position. For any orthogonal transformation $\mathbf Q$ and translation $\mathbf t$, the model satisfies

\begin{equation}
	\mathbf s\mapsto\mathbf s,\qquad
	\mathbf v\mapsto\mathbf v\mathbf Q^{\mathsf T},\qquad
	\Phi_\theta(\mathbf Q\mathbf r+\mathbf t;\{\mathbf Q\mathbf R+\mathbf t\})
	=\Phi_\theta(\mathbf r;\{\mathbf R\}).
\end{equation}

Continuous filters combine a radial Bessel basis with a compact $C^\infty$ envelope. We use the same basis for atom--atom (AA) and atom--grid (AG) edges, but shift only the AG distance to regularize grid points near a nucleus:

\begin{align}
	b_k^{x}(r)
	&=\sqrt{\frac{2}{R_c}}
	\frac{\sin\!\left[k\pi(r+\delta_x)/R_c\right]}{r+\delta_x},
	\quad k=1,\ldots,8,
	\delta_{\mathrm{AA}}=0,\quad \delta_{\mathrm{AG}}=0.5~\mathrm{Bohr}, \\
	c(r)&=
	\begin{cases}
		e\exp\!\left[-\dfrac{1}{1-(r/R_c)^2}\right],&r<R_c,\\
		0,&r\ge R_c,
	\end{cases}
	\qquad R_c=5~\text{\AA}.
\end{align}

The implementation evaluates the removable $r=0$ singularity through the corresponding sinc limit. Every multilayer perceptron uses shifted Softplus, $\operatorname{ssp}(x)=\ln(1+e^x)-\ln2$. The atom-centred integration grids contain no points exactly at a nucleus. On this domain, the radial filters, $c(r)$, and shifted Softplus support the derivatives required for density and density-gradient evaluation.

For an AA edge $n\rightarrow m$, let $\mathbf r_{mn}=\mathbf R_m-\mathbf R_n$, $r_{mn}=\|\mathbf r_{mn}\|$, and $\widehat{\mathbf r}_{mn}=\mathbf r_{mn}/(r_{mn}+\epsilon_r)$ with $\epsilon_r=10^{-8}~\mathrm{Bohr}$. A scalar network and a radial filter produce three channel-wise gates,

\begin{align}
	\mathbf z_{mn}^{\ell}
	&=\mathcal M_{\mathrm{AA}}^{\ell}(\mathbf s_n^\ell)
	\odot
	\left[\mathbf W_{\mathrm{AA}}^{\ell}\mathbf b^{\mathrm{AA}}(r_{mn})
	+\boldsymbol\eta_{\mathrm{AA}}^\ell\right]c(r_{mn})
	=\left[\mathbf z_{mn}^{s}\,\|\,\mathbf z_{mn}^{v}\,\|\,\mathbf z_{mn}^{r}\right], \\
	\Delta\mathbf s_m^{\mathrm{msg}}
	&=\sum_{n\in\mathcal N_m}\mathbf z_{mn}^{s}, \\
	\Delta\mathbf v_m^{\mathrm{msg}}
	&=\sum_{n\in\mathcal N_m}
	\left(\mathbf v_n^\ell\odot\mathbf z_{mn}^{v}
	+\mathbf z_{mn}^{r}\otimes\widehat{\mathbf r}_{mn}\right).
\end{align}

Here, $\|$ denotes concatenation, $\odot$ denotes channel-wise multiplication, and $\otimes$ broadcasts a scalar channel over a Cartesian direction. The first vector term transports directional features from the sender. The second couples an invariant gate to the edge direction and therefore creates an equivariant vector. Residual addition gives $\widetilde{\mathbf s}_m=\mathbf s_m^\ell+\Delta\mathbf s_m^{\mathrm{msg}}$ and $\widetilde{\mathbf v}_m=\mathbf v_m^\ell+\Delta\mathbf v_m^{\mathrm{msg}}$.

The subsequent self-update mixes scalar and vector channels through invariant norms and inner products. With bias-free channel maps $\mathbf U^\ell$ and $\mathbf V^\ell$, we define

\begin{align}
	\mathbf A_m&=\mathbf V^\ell\widetilde{\mathbf v}_m, \\
	\mathbf B_m&=\mathbf U^\ell\widetilde{\mathbf v}_m, \\
	\mathbf q_m&=\sqrt{\sum_{\alpha=1}^{3}\mathbf A_{m,:,\alpha}^{2}+\epsilon_v}, \\
	[\mathbf a_m\|\mathbf b_m\|\mathbf d_m]
	&=\mathcal M_{\mathrm{upd}}^\ell([\widetilde{\mathbf s}_m\|\mathbf q_m])
	\odot\tanh(\mathbf P^\ell\widetilde{\mathbf s}_m), \\
	\mathbf c_m&=\sum_{\alpha=1}^{3}
	\mathbf A_{m,:,\alpha}\odot\mathbf B_{m,:,\alpha}, \\
	\overline{\mathbf s}_m
	&=\widetilde{\mathbf s}_m+\mathbf a_m+\mathbf b_m\odot\mathbf c_m, \\
	\overline{\mathbf v}_m
	&=\widetilde{\mathbf v}_m+\mathbf d_m\odot\mathbf B_m.
\end{align}

We set the feature-norm stabilizer to $\epsilon_v=10^{-8}$. The $\tanh(\mathbf P^\ell\widetilde{\mathbf s}_m)$ factor makes the update vanish with the scalar state, while $\mathbf q_m$ and $\mathbf c_m$ preserve rotational invariance. The model caches $(\overline{\mathbf s}_m,\overline{\mathbf v}_m)$ for the grid stream before projecting to the next atom width. The projected atom state then undergoes an equivariant normalization,

\begin{align}
	\mathbf s_m^{\ell+1}
	&=\operatorname{LayerNorm}(\mathbf P_s^\ell\overline{\mathbf s}_m), \\
	\widehat{\mathbf v}_m&=\mathbf P_v^\ell\overline{\mathbf v}_m, \\
	\mathbf v_m^{\ell+1}
	&=\boldsymbol\gamma^\ell\odot
	\frac{\widehat{\mathbf v}_m}
	{\sqrt{\operatorname{mean}_{c,\alpha}(\widehat{\mathbf v}_{m,c,\alpha}^{2})+\epsilon_v}}.
\end{align}

The grid stream receives information only from the cached atom states. Independent AG networks compute $\mathbf z_{pi}^{\mathrm{AG}}$ with the same three-gate construction, using $\mathbf b^{\mathrm{AG}}(r_{pi})$ and the direction from atom $i$ to grid point $p$. The resulting messages are

\begin{align}
	\Delta\mathbf g_p^{\mathrm{msg}}
	&=\sum_{i\in\mathcal N_p}\mathbf z_{pi}^{s}, \\
	\Delta\mathbf u_p^{\mathrm{msg}}
	&=\sum_{i\in\mathcal N_p}
	\left(\overline{\mathbf v}_i\odot\mathbf z_{pi}^{v}
	+\mathbf z_{pi}^{r}\otimes\widehat{\mathbf r}_{pi}\right).
\end{align}

After adding the scalar message, an independent gated self-update $\mathcal U_{\mathrm G}^\ell$ acts on $(\mathbf g_p^\ell+\Delta\mathbf g_p^{\mathrm{msg}},\mathbf u_p^\ell)$. If it returns $(\Delta\mathbf g_p^{\mathrm{upd}},\Delta\mathbf u_p^{\mathrm{upd}})$, the layer output is

\begin{align}
	\mathbf g_p^{\ell+1}
	&=\mathbf P_g^\ell
	\left(\mathbf g_p^\ell+\Delta\mathbf g_p^{\mathrm{msg}}
	+\Delta\mathbf g_p^{\mathrm{upd}}\right), \\
	\mathbf u_p^{\ell+1}
	&=\mathbf P_u^\ell
	\left(\mathbf u_p^\ell+\Delta\mathbf u_p^{\mathrm{msg}}
	+\Delta\mathbf u_p^{\mathrm{upd}}\right).
\end{align}

The AG message functions, grid self-update, and grid projections use parameters independent of their atom counterparts. This separation lets the atom stream encode chemical environments while the grid stream represents the local field at query points. A bias-free scalar head finally produces

\begin{equation}
	\Phi_\theta(\mathbf r_p)=\mathbf w_2^{\mathsf T}
	\operatorname{ssp}(\mathbf W_1\mathbf g_p^L).
\end{equation}

The gated grid updates and bias-free grid projections preserve a zero grid state. Because $\operatorname{ssp}(0)=0$ and the readout also contains no bias, a grid point without incoming edges has $\Phi_\theta(\mathbf r_p)=0$. Element-specific heads map the final atom scalars to the coefficients and positive widths of the analytical Gaussian field described next.

\subsection{Singularity Mitigation and Boundary Anchoring}

Unlike PW-type data, LCAO data utilizes an all-electron basis set, which induces numerical spikes at the atomic nuclei, thereby hindering the neural network's ability to fit the data proficiently. To alleviate this, we employ a Delta-Learning approach to remove these singularities. We use a linear combination of Gaussian functions to fit the atomic electron density and determine the parameters of the Gaussian functions. Consequently, the model only needs to learn the residual difference between the true molecular electron density and the fitted one. This residual difference is considerably smoother and thus easier to learn.

DeepHartree learns the Hartree potential rather than the electron density directly. A Gaussian electron density corresponds analytically to a Hartree potential of the form $a\frac{\text{erf}(br)}{r}$, so the fitted density parameters directly construct the Hartree potential.

\begin{equation}
	V_H(\bm r) = \sum_{i\geq 0} \left( \underbrace{\sum_{k\geq 0} \gamma_{i,k}\frac{\text{erf}(\tau_{ik}|\bm r-\bm R_i|)}{|\bm r-\bm R_i|}}_{\text{Fixed Atomic Prior}} + \underbrace{\sum_{j\geq 0} \alpha_{i,j} \frac{\text{erf}(\beta_{ij} |\bm r - \bm R_i|)}{|\bm r - \bm R_i|}}_{\text{Learnable Environment}} \right) + \underbrace{ \Phi_{\theta}(\bm r,\{\bm R\})}_{\text{Grid ReadOut}}
\end{equation}

where $i$ indexes the atoms, $j$ indexes a set of basis functions with learnable coefficients $\alpha_{i,j}$ and widths $\beta_{ij}$ and $k$ indexes a set of basis functions with fixed coefficients $\gamma_{i,k}$ and widths $\tau_{ik}$. The term $\text{erf}(\beta r)/r$ behaves as $1/r$ in the long range but remains finite at the origin (converging to $2\beta/\sqrt{\pi}$), thereby removing the singularity. By summing over multiple widths (multi-scale basis), this baseline potential can flexibly fit the steep peaks near the nucleus and the smooth decay in the valence region, while the neural network term $\Phi(\bm r,\{R\})$ learns the remaining residual corrections.

The electron density follows the same analytical--local decomposition,

\begin{align}
	\rho(\bm r)&=\rho_{\mathrm G}(\bm r)+\rho_{\theta}(\bm r), \\
	\rho_{\mathrm G}(\bm r)
	&=\sum_{i,k}\gamma_{i,k}\frac{\tau_{ik}^{3}}{\pi^{3/2}}
	\exp\!\left[-\tau_{ik}^{2}|\bm r-\bm R_i|^{2}\right]
	+\sum_{i,j}\alpha_{i,j}\frac{\beta_{ij}^{3}}{\pi^{3/2}}
	\exp\!\left[-\beta_{ij}^{2}|\bm r-\bm R_i|^{2}\right], \\
	\rho_{\theta}(\bm r)&=-\frac{1}{4\pi}\nabla^{2}\Phi_{\theta}(\bm r,\{\bm R\}).
\end{align}

We evaluate the fixed and learnable Gaussian terms directly from their closed-form expressions, without differentiating their Hartree potentials. Automatic differentiation is applied only to the learned local potential $\Phi_\theta$ to obtain the local density $\rho_\theta$.

This decomposition combines long-range electrostatics with efficient local message passing. The atom-parameterized erf terms carry the Hartree potential to distant grid points, while the grid readout learns local corrections from a fixed number of nearby atoms. The resulting semi-analytical representation preserves the long-range potential and resolves environment-dependent structure without dense atom--grid connectivity.

Furthermore, the exact $1/r$ asymptotic decay of this analytical prior, synergizing with the zero-bias constraint and spatial cutoff of the grid readout network $\Phi_{\theta}$, ensures that the total predicted potential strictly vanishes at infinity ($\lim_{|\bm r| \to \infty} V_H(\bm r) = 0$). This inherent satisfaction of the physical boundary condition perfectly anchors the gauge freedom of the Poisson equation, thereby preventing any systematic global drift in the predicted Hartree potential.

\subsection{Training details}

We use the same training pipeline for the QM9 and VQM24 models. The loader preloads molecular metadata and stored AO density matrices from HDF5 shards and constructs each PySCF molecule object once. When drawing a sample, it reconstructs the full symmetric density matrix from upper-triangular storage. We reserve 10\% of each dataset for testing and draw 100 validation molecules from the remainder. We save the split identifiers for reproducibility. Training uses shuffled minibatches of 64 molecules, whereas validation and test process one molecule at a time.

PySCF constructs a level-0 atom-centred quadrature grid whenever the loader draws a molecule. During training, we sample up to 100 grid points per molecule uniformly without replacement; validation and test use every grid point. We evaluate density targets from the stored AO density matrices. The loader precomputes bidirectional AA edges, while the batch collation constructs AG edges with the same $5~\text{\AA}$ cutoff used by the model. At each sampled point, the predicted density combines the analytical Gaussian contribution with the automatic-differentiation Laplacian of the local potential $\Phi_\theta$.

\par\begingroup
We minimize a quadrature-weighted $L_1$ loss averaged over the molecules in each minibatch $\mathcal B$,

\begin{equation}
\mathcal L
=\frac{1}{|\mathcal B|}\sum_{i\in\mathcal B}
\frac{N_i}{n_i}\sum_{p\in S_i}
|w_{ip}|\,|\hat\rho_{ip}-\rho_{ip}|,
\end{equation}

where $N_i$ is the full grid size, $S_i$ contains the $n_i$ sampled points, and $w_{ip}$ is the quadrature weight. The factor $N_i/n_i$ corrects for uniform subsampling of each molecular grid.
\par\endgroup

We train for at most $10^6$ optimization steps using AdamW with an initial learning rate of $10^{-3}$, $(\beta_1,\beta_2)=(0.9,0.999)$, $\epsilon=10^{-8}$, and weight decay $10^{-2}$. We clip the global gradient norm at 1.0. The learning-rate multiplier uses a 1,000-step linear warmup followed by exponential power decay,

\begin{equation}
\lambda(t)=
\begin{cases}
(t+1)/1000, & t<1000,\\
0.96^{(t-1000)/10^4}, & t\geq1000,
\end{cases}
\qquad \eta_t=10^{-3}\lambda(t).
\end{equation}

\par\begingroup
Every 1,000 steps, we evaluate the mean molecular integrated $L_1$ error on the complete validation grids, with $L_{1,i}=\sum_p w_{ip}|\hat\rho_{ip}-\rho_{ip}|$. We save checkpoints every 5,000 steps and use the model weights with the lowest validation $L_1$ error for testing. Training stops if this error does not improve for 100,000 steps.
\par\endgroup

\subsection{One-Shot Construction of the Kohn--Sham Density Matrix}

DeepHartree maps its predicted real-space fields to an atomic-orbital (AO) density matrix through one non-self-consistent Kohn--Sham build and generalized diagonalization. For a molecular quadrature grid $\{\bm r_g,w_g\}_{g=1}^{N_G}$, the model supplies $\widehat V_H(\bm r_g)$, $\widehat\rho(\bm r_g)$, and $\nabla\widehat\rho(\bm r_g)$. We combine these fields with analytical one-electron integrals to assemble the effective Kohn--Sham matrix. The resulting orbitals support one-shot property evaluation, and the density matrix provides an initial guess for a conventional self-consistent-field (SCF) calculation. {The QM9 and VQM24 one-shot benchmarks use spin-unpolarized PBE. AD constructs the matrix on PySCF level-3 atom-centred grids. Hybrid combines analytical Gaussian Coulomb integrals with level-0 quadrature for the local Hartree and full-density XC terms, plus a level-3 Gaussian-density XC correction (Eq.~\ref{eq:hybrid_effective_potential}). For both methods, post-diagonalization XC energies are evaluated from the resulting density matrices on level-3 grids.}

The field-dependent Kohn--Sham matrix takes the form

\begin{equation}
	\widehat{\mathbf F}_{\mathrm{KS}}
	=\mathbf H_{\mathrm{core}}
	+\widehat{\mathbf J}
	+\widehat{\mathbf V}_{\mathrm{xc}},
	\qquad
	\mathbf H_{\mathrm{core}}=\mathbf T+\mathbf V_{\mathrm{ne}},
	\label{eq:one_shot_ks_matrix}
\end{equation}

where $\mathbf T$ and $\mathbf V_{\mathrm{ne}}$ denote the kinetic-energy and electron--nuclear-attraction matrices. PySCF evaluates both matrices analytically in the chosen AO basis. The predicted Hartree potential defines the Coulomb contribution as a local multiplicative operator,

\begin{align}
	\widehat J_{\mu\nu}
	&=\int \chi_\mu(\bm r)\widehat V_H(\bm r)\chi_\nu(\bm r)\,\mathrm d\bm r \\
	&\approx\sum_{g=1}^{N_G}w_g\,
	\chi_\mu(\bm r_g)\widehat V_H(\bm r_g)\chi_\nu(\bm r_g).
	\label{eq:one_shot_coulomb}
\end{align}

For the spin-unpolarized generalized-gradient approximation (GGA) considered here, we write the exchange--correlation (XC) functional as

\begin{equation}
	E_{\mathrm{xc}}[\widehat\rho]
	=\int f_{\mathrm{xc}}\!\left(\widehat\rho(\bm r),\widehat\sigma(\bm r)\right)\,\mathrm d\bm r,
	\qquad
	\widehat\sigma(\bm r)=|\nabla\widehat\rho(\bm r)|^2.
	\label{eq:gga_functional}
\end{equation}

Here, $f_{\mathrm{xc}}$ denotes the XC energy density per unit volume. We define $v_\rho=\partial f_{\mathrm{xc}}/\partial\rho$ and $v_\sigma=\partial f_{\mathrm{xc}}/\partial\sigma$, evaluated at $(\widehat\rho,\widehat\sigma)$. Functional differentiation gives $\delta E_{\mathrm{xc}}/\delta\rho=v_\rho-2\nabla\!\cdot(v_\sigma\nabla\widehat\rho)$. Integration by parts then yields the AO matrix elements

\begin{align}
	[\widehat{\mathbf V}_{\mathrm{xc}}]_{\mu\nu}
	&=\int \chi_\mu\chi_\nu
	\frac{\delta E_{\mathrm{xc}}}{\delta\rho}\,\mathrm d\bm r \\
	&=\int\!\left[
	v_\rho\chi_\mu\chi_\nu
	+2v_\sigma\nabla\widehat\rho\cdot
	\left(\nabla\chi_\mu\,\chi_\nu+\chi_\mu\nabla\chi_\nu\right)
	\right]\mathrm d\bm r \\
	&\approx\sum_{g=1}^{N_G}w_g\left[
	v_{\rho,g}\chi_{\mu,g}\chi_{\nu,g}
	+2v_{\sigma,g}\nabla\widehat\rho_g\cdot
	\left(\nabla\chi_{\mu,g}\chi_{\nu,g}
	+\chi_{\mu,g}\nabla\chi_{\nu,g}\right)
	\right],
	\label{eq:one_shot_xc}
\end{align}

where a subscript $g$ denotes evaluation at $\bm r_g$, for example $\chi_{\mu,g}=\chi_\mu(\bm r_g)$. For an isolated molecule, the Gaussian AOs and electron density decay at infinity; provided that $v_\sigma\nabla\widehat\rho$ remains bounded, the corresponding surface term vanishes. The resulting weak form transfers the divergence operator to the AO product. It therefore requires only the model-provided first derivative $\nabla\widehat\rho$ and analytical AO gradients; neither $\nabla^2\widehat\rho$ nor $\nabla v_\sigma$ is evaluated on the molecular grid. We obtain $v_\rho$ and $v_\sigma$ from the Libxc interface in PySCF and GPU4PySCF \citep{perdew1996generalized,lehtola2018recent,sun2018pyscf,sun2020recent,li2025gpu4pyscf}.

We obtain the molecular orbitals from one solution of the Roothaan--Hall generalized eigenvalue problem,

\begin{equation}
	\widehat{\mathbf F}_{\mathrm{KS}}\mathbf C
	=\mathbf S\mathbf C\boldsymbol\varepsilon,
	\qquad
	\mathbf C^{\mathsf T}\mathbf S\mathbf C=\mathbf I,
	\label{eq:one_shot_roothaan_hall}
\end{equation}

where $\mathbf S$ is the AO overlap matrix, $\mathbf C$ contains the molecular-orbital coefficients, and $\boldsymbol\varepsilon$ is the diagonal orbital-energy matrix. For the restricted closed-shell systems considered here, $N_{\mathrm{occ}}=N_{\mathrm e}/2$. Occupation of the lowest $N_{\mathrm{occ}}$ orbitals gives

\begin{equation}
	\mathbf D^{(0)}
	=2\mathbf C_{\mathrm{occ}}\mathbf C_{\mathrm{occ}}^{\mathsf T},
	\qquad
	\operatorname{Tr}\!\left[\mathbf D^{(0)}\mathbf S\right]
	=2N_{\mathrm{occ}}=N_{\mathrm e}.
	\label{eq:one_shot_density_matrix}
\end{equation}

Here, $\mathbf C_{\mathrm{occ}}$ collects the occupied columns of $\mathbf C$. The learned fields determine $\widehat{\mathbf F}_{\mathrm{KS}}$ before diagonalization, so the one-shot construction contains no fixed-point iteration. We use $\mathbf D^{(0)}$, together with the orbital energies and coefficients from the same diagonalization, for the reported one-shot properties. For SCF acceleration, we pass $\mathbf D^{(0)}$ to the standard SCF solver as the initial density matrix.

% !TEX root = ../main.tex
\section{Conclusion}

DeepHartree establishes a direct interface between learned real-space fields and LCAO electronic structure. The network predicts a continuous Hartree potential, and the Poisson equation fixes its relation to the electron density. Atom-centred Gaussian fields resolve the near-nuclear structure, while charge correction fixes the molecular electron count. {Analytical Gaussian integrals and numerical quadrature assemble the Kohn--Sham matrix in the target AO basis, and one diagonalization yields the corresponding density matrix.} This separation lets the model learn a physical field instead of the dimensions and conventions of one matrix representation.

The density benchmarks show that this field representation preserves information required beyond grid-level regression. {Density benchmarks quantify accuracy on both coarse and fine grids, while ablations establish the contributions of fixed Gaussian priors, learnable shells, and charge conservation.} {A single diagonalization provides estimates of frontier orbital energies and the occupied subspace. Its density matrix supports energy-component evaluation, while its orbitals allow multi-level spectral comparisons.} {Across QM9 and VQM24, models trained on def2-SVP densities retain total-energy accuracy and occupied-subspace agreement when transferred to def2-TZVP and def2-QZVP without retraining. On QM9, AD total-energy MAEs remain within 10.63--10.89~meV atom$^{-1}$ across the three bases. Together, these results connect the model's physical structure to transferable electronic-structure reconstruction.}

{Hybrid} turns the same representation into a scalable inference route. The method uses a seven-point stencil to obtain the learned local density and first-order automatic differentiation to obtain its gradient. It removes the higher-order graphs that make full AD slower and numerically unstable on large molecular grids. {Combined with analytical Gaussian integrals and multilevel quadrature, this approach retains total-energy accuracy across the tested bases, with property-dependent changes in frontier orbital errors. Excluding grid construction, Hybrid achieves near-linear density-matrix construction in both molecular families. For the 1004-atom PEG chain, it is 8.75-fold faster than AD and 7.34-fold faster than GPU4PySCF.} {Without fine-tuning, AD evaluation of the QM9-trained model yields a total-energy MAE of 9.437~meV atom$^{-1}$ on OE62 molecules containing 30--106 atoms. Component-wise accuracy is retained under size extrapolation, with error cancellation supporting the total-energy accuracy.}

The predicted matrices also improve self-consistent calculations. {DeepHartree initialization reduces mean QM9 SCF iteration counts by 15.3--15.6\% across the three basis sets, with reductions for 96.8--97.5\% of comparable molecules. For the 620-atom organic cage, the reduction in SCF time exceeds the initialization cost, lowering total wall-clock time by 12.9\%. These results connect transferable one-shot density matrices to faster self-consistent calculations.} {Future extensions can build on this field-to-LCAO framework to address charged and open-shell systems, periodic structures, and higher-rung density functionals.} {DeepHartree therefore links physically constrained neural fields to transferable electronic-structure reconstruction and accelerated SCF workflows.}

% !TEX root = ../main.tex
\section{Dataset}

We constructed training and test density-matrix datasets from QM9 and VQM24. QM9 contains approximately 134,000 equilibrium organic molecules formed from H, C, N, O, and F \citep{ramakrishnan2014quantum}. We retained 133,885 neutral closed-shell structures. VQM24 enumerates neutral closed-shell organic and inorganic molecules with up to five heavy atoms from C, N, O, F, Si, P, S, Cl, and Br \citep{khan2025vqm24}. We used the 258,242 structures in \texttt{DFT\_uniques.npz}, which contains the lowest-energy conformer for each constitutional isomer.

We recomputed the reference density matrices for both datasets at the PBE/def2-SVP level with density fitting in PySCF and GPU4PySCF \citep{sun2018pyscf,sun2020recent,li2025gpu4pyscf}. This common calculation level separates the present density-learning targets from the quantum-chemical properties distributed with the source datasets. Each HDF5 record stores the converged upper-triangular density matrix, atomic coordinates, elements, and calculation metadata. Storing density matrices avoids a fixed Cartesian box and lets us evaluate the density on any later integration grid.

Training used PySCF level-0 atom-centred grids and sampled at most 100 grid points per molecule in each update. Validation and test evaluations used every point on the corresponding grids. The fixed QM9 split contains 120,396 training, 100 validation, and 13,389 test molecules. The VQM24 split contains 232,317 training, 100 validation, and 25,825 test molecules. We report molecular distributions without pooling the two test sets because their element compositions and density scales differ.

OE62 provides 61,489 unique organic molecules extracted from experimentally reported molecular crystals and relaxed in the gas phase with DFT \citep{stuke2020atomic}. We used OE62 only for external size-OOD evaluation; no OE62 structure entered model training, validation, or checkpoint selection. For the selected geometries, we used stored PBE/def2-SVP reference density matrices generated with density fitting and evaluated each matrix once without a new SCF calculation. We retained H/C/N/O/F molecules with more than 29 atoms and selected 1,000 structures through proportional exact-atom-count stratification. The resulting set spans 30--106 atoms and 71 represented atom counts.

% !TEX root = ../main.tex
\section*{Code availability}
We provide the DeepHartree source code and the scripts for dataset preparation, training, and evaluation at \url{https://github.com/presentjjjjk/DeepHartree-for-DFT}.

\section*{Data availability}
We provide the QM9 and VQM24 PBE/def2-SVP density-matrix datasets at \url{https://huggingface.co/datasets/jiankunwu/DeepHartree-PBE-def2SVP-density-matrices}.

% !TEX root = ../main.tex
\section*{Acknowledgements}
Generous financial support by the National Natural Science Foundation of China (U24A20527) and the Key Research and Development Program of Zhejiang Province (2023C01102).

\bibliography{ref}

\clearpage
\appendix
\setcounter{secnumdepth}{1}
\renewcommand{\thesection}{Appendix~\Alph{section}}
\numberwithin{equation}{section}
\counterwithin{table}{section}
\counterwithin{figure}{section}
\renewcommand{\theequation}{\Alph{section}\arabic{equation}}
\renewcommand{\thetable}{\Alph{section}\arabic{table}}
\renewcommand{\thefigure}{\Alph{section}\arabic{figure}}
% Keep PDF anchors unique after resetting appendix counters.
\renewcommand{\theHequation}{appendix.\Alph{section}.\arabic{equation}}
\renewcommand{\theHtable}{appendix.\Alph{section}.\arabic{table}}
\renewcommand{\theHfigure}{appendix.\Alph{section}.\arabic{figure}}
% !TEX root = ../main.tex
% Adapted from Supporting_Infomation/main.tex; scientific content retained.
\section{Training configuration}
\label{app:training}

Table~\ref{tab:training-settings} summarizes the hyperparameters required to reproduce the QM9 and VQM24 model training.

\begin{table}[htbp]
  \centering
  \caption{Hyperparameters used to train the QM9 and VQM24 models.}
  \label{tab:training-settings}
  \small
  \begin{tabularx}{\linewidth}{@{}L{0.35\linewidth}Y@{}}
    \toprule
    Setting & Value \\
    \midrule
    Atomic embedding & 64 entries with 64-dimensional embeddings \\
    PaiNN channel widths & $64\rightarrow128\rightarrow128\rightarrow64$ \\
    Radial basis & Eight functions; $5~\text{\AA}$ cutoff; 0.5~Bohr atom--grid distance shift \\
    Grid-potential head & $64\rightarrow64\rightarrow1$ bias-free projection with shifted-Softplus activation \\
    Learnable Gaussian shells & QM9, H: 4 and C/N/O/F: 16; VQM24, H: 4, C/N/O/F: 16, Si/P/S/Cl: 24, and Br: 32 \\
    Minimum learnable Gaussian width & $10^{-3}$ Bohr \\
    Training grid and sampling & PySCF level-0 grid; at most 100 points per molecule, sampled uniformly without replacement \\
    Loss & Molecular mean of the absolute density error weighted by $|w_p|$, corrected by the full-to-sampled grid-size ratio \\
    Minibatch size & 64 molecules \\
    Optimizer & AdamW; initial learning rate $10^{-3}$; weight decay $10^{-2}$; gradient-norm limit 1.0 \\
    Learning-rate schedule & 1,000-step linear warmup followed by $0.96^{(t-1000)/10^4}$ decay \\
    Training and selection & At most $10^6$ steps; lowest mean molecular integrated $L_1$ error on 100 validation molecules, evaluated using signed quadrature weights $w_p$ \\
    \bottomrule
  \end{tabularx}
\end{table}

\clearpage
\section{Fixed-density fitting}
\label{app:fixed-density}

The fixed density represents the spherically averaged density of a neutral isolated atom. Each reference was calculated with unrestricted Kohn--Sham PBE/def2-SVP in PySCF on a level-3 atom-centred grid. For element $Z$, we fit $K_Z$ normalized isotropic Gaussians,

\begin{equation}
  \rho_Z^{\mathrm{fix}}(r)
  =\sum_{k=1}^{K_Z}
  \frac{Q_{Zk}}{\pi^{3/2}\sigma_{Zk}^{3}}
  \exp\!\left(-\frac{r^2}{\sigma_{Zk}^{2}}\right),
  \qquad Q_{Zk}>0,\quad \sigma_{Zk}>0,\quad
  \sum_{k=1}^{K_Z}Q_{Zk}=Z.
  \label{eq:fixed-density}
\end{equation}
Here, $Q_{Zk}$ is the electron count assigned to component $k$, and $\sigma_{Zk}$ is its width in Bohr. The corresponding atom-centred Hartree potential has the closed form
\begin{equation}
  V_Z^{\mathrm{fix}}(r)
  =\sum_{k=1}^{K_Z}Q_{Zk}\frac{\operatorname{erf}(r/\sigma_{Zk})}{r},
  \label{eq:fixed-potential}
\end{equation}
with the finite $r\rightarrow0$ limit $2Q_{Zk}/(\sqrt{\pi}\sigma_{Zk})$ for each component. This parameterization corresponds to $\gamma_{ik}=Q_{Zk}$ and $\tau_{ik}=1/\sigma_{Zk}$ in the main-text notation. We fit the spherical density by minimizing a quadrature-weighted squared error. Positive widths are enforced by exponentiation, while a softmax enforces positive charges and $\sum_kQ_{Zk}=Z$.

The QM9 fits use $K_{\mathrm H}=4$ and $K_Z=12$ for C, N, O, and F. Table~\ref{tab:qm9-fixed-parameters} reports the resulting parameters. Values are rounded to three decimal places for presentation; the implementation reads the full-precision values from the accompanying JSON files. Each $Q$ list and $\sigma$ list follows the same component order.

\begingroup
\small
\setlength{\tabcolsep}{4pt}
\begin{longtable}{@{}L{0.08\linewidth}L{0.43\linewidth}L{0.43\linewidth}@{}}
  \caption{QM9 fixed-density Gaussian parameters, rounded to three decimal places. $Q_k$ is in electrons and $\sigma_k$ is in Bohr.}
  \label{tab:qm9-fixed-parameters}\\
  \toprule
  Element & $\{Q_k\}$ & $\{\sigma_k\}$ \\
  \midrule
  \endfirsthead
  \multicolumn{3}{c}{\tablename~\thetable\ (continued)}\\
  \toprule
  Element & $\{Q_k\}$ & $\{\sigma_k\}$ \\
  \midrule
  \endhead
  \midrule
  \multicolumn{3}{r}{Continued on next page}\\
  \endfoot
  \bottomrule
  \endlastfoot
  H & 0.027, 0.328, 0.318, 0.327 & 0.350, 1.408, 0.820, 1.791 \\
  C & 0.029, 0.342, 1.042, 0.719, 0.579, 0.525, 0.494, 0.475, 0.456, 0.444, 0.438, 0.456 & 0.051, 0.115, 0.210, 1.323, 1.404, 1.343, 1.137, 0.330, 1.109, 1.110, 2.070, 2.132 \\
  N & 0.030, 0.357, 1.083, 0.933, 0.763, 0.668, 0.639, 0.586, 0.543, 0.406, 0.502, 0.490 & 0.044, 0.099, 0.182, 1.127, 1.154, 0.984, 1.742, 1.739, 0.792, 0.289, 0.961, 1.075 \\
  O & 0.032, 0.375, 1.126, 1.195, 0.953, 0.824, 0.759, 0.688, 0.613, 0.343, 0.550, 0.541 & 0.039, 0.087, 0.161, 0.892, 0.892, 1.361, 1.417, 1.410, 0.676, 0.265, 0.811, 0.810 \\
  F & 0.034, 0.414, 0.887, 1.188, 1.036, 0.893, 0.796, 0.740, 0.689, 1.232, 0.426, 0.665 & 0.035, 0.080, 0.736, 0.755, 0.760, 1.306, 1.195, 0.769, 0.768, 0.150, 0.354, 1.415 \\
\end{longtable}
\endgroup

The VQM24 fits use $K_{\mathrm H}=4$, $K_Z=12$ for C, N, O, and F, $K_Z=16$ for Si, P, S, and Cl, and $K_{\mathrm{Br}}=24$. Table~\ref{tab:vqm24-fixed-parameters} lists the fitted parameters.

\begingroup
\footnotesize
\setlength{\tabcolsep}{4pt}
\begin{longtable}{@{}L{0.08\linewidth}L{0.43\linewidth}L{0.43\linewidth}@{}}
  \caption{VQM24 fixed-density Gaussian parameters, rounded to three decimal places. $Q_k$ is in electrons and $\sigma_k$ is in Bohr.}
  \label{tab:vqm24-fixed-parameters}\\
  \toprule
  Element & $\{Q_k\}$ & $\{\sigma_k\}$ \\
  \midrule
  \endfirsthead
  \multicolumn{3}{c}{\tablename~\thetable\ (continued)}\\
  \toprule
  Element & $\{Q_k\}$ & $\{\sigma_k\}$ \\
  \midrule
  \endhead
  \midrule
  \multicolumn{3}{r}{Continued on next page}\\
  \endfoot
  \bottomrule
  \endlastfoot
  H & 0.027, 0.328, 0.318, 0.326 & 0.350, 1.409, 0.820, 1.792 \\
  C & 0.040, 0.480, 0.718, 0.603, 0.504, 0.455, 0.424, 0.389, 0.385, 0.366, 1.278, 0.357 & 0.055, 0.127, 0.839, 1.419, 1.419, 1.421, 1.423, 2.439, 1.423, 1.422, 0.242, 1.422 \\
  N & 0.039, 0.476, 1.218, 0.803, 0.602, 0.516, 0.453, 0.435, 0.417, 1.269, 0.394, 0.378 & 0.047, 0.108, 0.748, 1.223, 1.223, 1.223, 2.106, 1.224, 1.224, 0.205, 1.224, 1.224 \\
  O & 0.004, 0.080, 2.712, 0.911, 0.589, 0.479, 0.429, 0.388, 0.355, 0.343, 1.147, 0.563 & 0.022, 0.053, 0.823, 1.092, 1.148, 1.483, 1.714, 1.531, 0.971, 0.520, 0.186, 0.105 \\
  F & 0.004, 0.083, 2.630, 1.223, 0.812, 0.628, 0.517, 0.472, 0.576, 1.138, 0.459, 0.458 & 0.019, 0.048, 0.626, 1.341, 0.875, 0.948, 0.980, 0.946, 0.094, 0.166, 1.128, 1.128 \\
  Si & 0.042, 0.450, 0.915, 1.535, 1.354, 0.964, 1.181, 0.753, 0.735, 0.767, 0.775, 0.887, 0.909, 0.911, 0.909, 0.914 & 0.023, 0.052, 0.407, 0.493, 0.432, 0.286, 0.095, 0.404, 0.540, 0.567, 0.386, 1.851, 1.828, 1.791, 1.755, 2.906 \\
  P & 0.043, 0.459, 1.090, 1.687, 1.318, 1.077, 0.897, 1.169, 0.772, 0.821, 0.919, 0.945, 0.950, 0.949, 0.951, 0.952 & 0.022, 0.049, 0.389, 0.414, 0.277, 0.431, 0.454, 0.089, 0.456, 0.350, 1.616, 1.615, 1.611, 1.608, 1.606, 2.610 \\
  S & 0.028, 0.368, 1.275, 2.049, 1.527, 1.102, 0.884, 1.207, 0.736, 0.923, 0.959, 0.983, 0.989, 0.991, 0.991, 0.987 & 0.018, 0.042, 0.379, 0.363, 0.365, 0.360, 0.345, 0.079, 0.210, 1.449, 1.451, 1.450, 1.448, 1.447, 1.446, 2.434 \\
  Cl & 0.028, 0.365, 1.127, 2.132, 1.520, 1.109, 0.899, 0.754, 0.992, 1.202, 1.085, 1.139, 1.156, 1.163, 1.165, 1.164 & 0.017, 0.039, 0.335, 0.328, 0.334, 0.339, 0.335, 0.198, 1.332, 0.074, 2.217, 1.332, 1.331, 1.332, 1.332, 1.332 \\
  Br & 0.024, 6.286, 0.345, 0.382, 1.200, 0.892, 1.140, 1.191, 1.373, 1.420, 1.441, 1.442, 1.435, 1.427, 1.414, 1.400, 1.390, 1.383, 1.377, 1.364, 1.638, 1.667, 1.681, 1.688 & 0.008, 0.125, 0.018, 0.125, 0.035, 0.446, 1.357, 1.331, 0.444, 0.444, 0.444, 0.444, 0.445, 0.445, 0.445, 0.445, 0.445, 0.446, 0.447, 0.450, 2.242, 1.509, 1.396, 1.398 \\
\end{longtable}
\endgroup

\clearpage

\section{Dipole moment from implicit atomic charges}
\label{app:dipole}

DeepHartree decomposes the electron density as $\rho=\rho_{\mathrm G}+\rho_\theta$, where $\rho_\theta=-\nabla^2\Phi_\theta/(4\pi)$. The local potential $\Phi_\theta$ is zero outside the union of the atomic cutoff regions. Its gradient also vanishes at the boundary because the radial envelope is smooth. For Cartesian component $a$, the identity
\begin{equation}
  \nabla\!\cdot\!\left(r_a\nabla\Phi_\theta-\Phi_\theta\mathbf e_a\right)
  =r_a\nabla^2\Phi_\theta
\end{equation}
and the divergence theorem therefore give
\begin{equation}
  \int_{\mathbb R^3}r_a\rho_\theta(\mathbf r)\,\mathrm d^3\mathbf r
  =-\frac{1}{4\pi}\lim_{R\to\infty}
  \oint_{S_R}\left(r_a\nabla\Phi_\theta-\Phi_\theta\mathbf e_a\right)
  \!\cdot\mathrm d\mathbf S
  =0.
\end{equation}
Thus, the local neural field contributes neither net electron number nor an electronic dipole moment.

Each Gaussian $G_{ik}$ is normalized, spherically symmetric, and centred at atom $i$. Its first moment is therefore
\begin{equation}
  \int_{\mathbb R^3}\mathbf r\,q_{ik}G_{ik}(\mathbf r-\mathbf R_i)\,
  \mathrm d^3\mathbf r=q_{ik}\mathbf R_i.
\end{equation}
Let $n_i=\sum_k q_{ik}$ include the fixed and learned Gaussian-shell charges on atom $i$. In atomic units, the molecular dipole is
\begin{equation}
  \boldsymbol\mu
  =\sum_i Z_i\mathbf R_i-\int_{\mathbb R^3}\mathbf r\,\rho(\mathbf r)\,\mathrm d^3\mathbf r
  =\sum_i\left(Z_i-n_i\right)\mathbf R_i
  =\sum_i\widetilde q_i\mathbf R_i,
\end{equation}
where $\widetilde q_i=Z_i-n_i$ is the model-derived implicit atomic charge. Charge conservation gives $\sum_i\widetilde q_i=0$ for the neutral molecules considered here, so the dipole is independent of the coordinate origin. Hence, DeepHartree evaluates the dipole directly from the implicit atomic charges and nuclear positions, without integrating the local density on a numerical grid.

\clearpage
\section{Hybrid finite-difference step-size scan}
\label{app:hybrid-step}

We compared Hybrid densities with full automatic-differentiation (AD) densities for the same trained QM9 model on a PySCF level-3 grid. A fixed random sample of 100 molecules from the 13,389-molecule QM9 test set was evaluated at ten steps from 0.01 to 0.10~Bohr in increments of 0.01~Bohr. For molecule $i$ and step $h$, the selection metric was
\begin{equation}
  \mathrm{wNMAE}_i(h)=
  \frac{\sum_p w_{ip}\left|\rho^{\mathrm{Hybrid}}_{i,h,p}-\rho^{\mathrm{AD}}_{i,p}\right|}
       {\sum_p w_{ip}\left|\rho^{\mathrm{AD}}_{i,p}\right|},
\end{equation}
where $w_{ip}$ are the signed PySCF quadrature weights. We selected the step with the lowest arithmetic mean of $\mathrm{wNMAE}_i(h)$ across the 100 molecules; neither $L_1$ nor $L_2$ entered the selection criterion.

\begin{figure}[htbp]
  \centering
  \includegraphics[width=\linewidth]{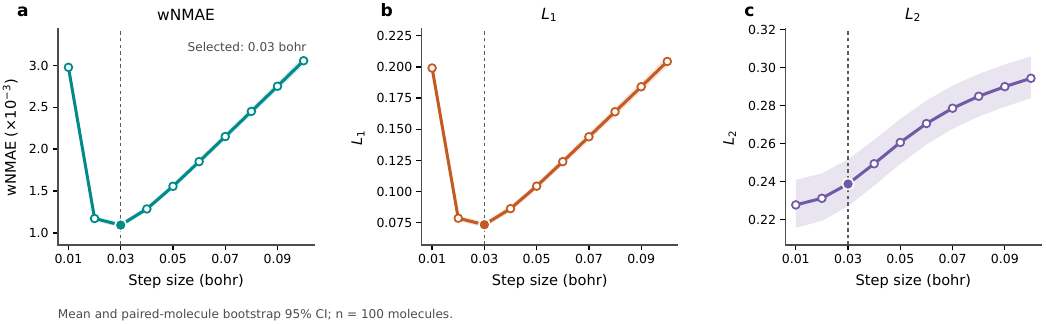}
  \caption{Hybrid step-size scan against the full-AD density on 100 QM9 test molecules. (a) Mean wNMAE, (b) mean $L_1$ error in electrons, and (c) mean $L_2$ error in $\mathrm{e}\,\mathrm{Bohr}^{-3/2}$. Shading denotes paired-molecule bootstrap 95\% confidence intervals for the means (10,000 resamples). The dashed line marks the selected 0.03~Bohr step.}
  \label{fig:hybrid-step-scan}
\end{figure}

The minimum mean wNMAE occurs at 0.03~Bohr (0.109\%), compared with 0.117\% at 0.02~Bohr and 0.129\% at 0.04~Bohr (Figure~\ref{fig:hybrid-step-scan}a). The same step also minimizes mean $L_1$ (0.073~e). By contrast, mean $L_2$ is smallest at 0.01~Bohr (0.228~$\mathrm{e}\,\mathrm{Bohr}^{-3/2}$), versus 0.239~$\mathrm{e}\,\mathrm{Bohr}^{-3/2}$ at the selected step. Thus, 0.03~Bohr is optimal among the scanned values for the prespecified wNMAE criterion, not for every error measure. This scan tests agreement with the model's AD density, rather than accuracy against a DFT reference density.

\clearpage
\section{One-shot energy and energy-component evaluation}
\label{app:energy}

The model-derived Kohn--Sham matrix is diagonalized once to obtain the occupied orbitals and the closed-shell AO density matrix $\mathbf D^{(0)}=2\mathbf C_{\mathrm{occ}}\mathbf C_{\mathrm{occ}}^{\mathsf T}$. Reported orbital energies come from this first diagonalization. The reported total energy and its components are instead evaluated from $\mathbf D^{(0)}$ using the same PBE/def2-SVP ,PBE/def2-TZVP or PBE/def2-QZVP convention as the corresponding reference. In atomic units, they are
\begin{align}
  E_{\mathrm{kin}}&=\operatorname{Tr}\!\left[\mathbf D^{(0)}\mathbf T\right],
  &E_{\mathrm{ne}}&=\operatorname{Tr}\!\left[\mathbf D^{(0)}\mathbf V_{\mathrm{ne}}\right],\\
  E_{\mathrm J}&=\tfrac12\operatorname{Tr}\!\left[\mathbf D^{(0)}\mathbf J[\mathbf D^{(0)}]\right],
  &E_{\mathrm{xc}}&=\int f_{\mathrm{xc}}\!\left(\rho[\mathbf D^{(0)}],
    |\nabla\rho[\mathbf D^{(0)}]|^2\right)\,\mathrm d^3\mathbf r,\\
  E_{\mathrm{nn}}&=\sum_{A<B}\frac{Z_AZ_B}{|\mathbf R_A-\mathbf R_B|},
  &E_{\mathrm{tot}}&=E_{\mathrm{kin}}+E_{\mathrm{ne}}+E_{\mathrm J}
    +E_{\mathrm{xc}}+E_{\mathrm{nn}}.
\end{align}
Here, $\mathbf T$ and $\mathbf V_{\mathrm{ne}}$ are the AO kinetic and electron--nuclear matrices. The Coulomb matrix $\mathbf J[\mathbf D^{(0)}]$ is rebuilt with density fitting, and $E_{\mathrm{xc}}$ is evaluated from the density of $\mathbf D^{(0)}$ on a PySCF level-3 grid. These post-diagonalization definitions are the same for AD and Hybrid; the predicted field affects them through $\mathbf D^{(0)}$. No further self-consistent iteration is performed.

\clearpage
\section{Large-system field evaluation and spatial screening}
\label{app:screening}

For large-system AD evaluation, each fixed or learnable Gaussian shell has charge $q_{ik}$ and width $\sigma_{ik}$. Writing $r_i=|\mathbf r-\mathbf R_i|$ and $Q_i=\sum_k q_{ik}$, its total Hartree potential is separated into a point-charge term and a localized correction:
\begin{equation}
  V_{\mathrm G}(\mathbf r)
  =\sum_i\frac{Q_i}{r_i}
  -\sum_{i,k}q_{ik}\frac{\operatorname{erfc}(r_i/\sigma_{ik})}{r_i}.
\end{equation}
For at most 400 Gaussian shells, the numerical AD path sums both terms directly on the GPU. Above 400 shells, FMM3D evaluates the atom-level point-charge term with a requested tolerance of $10^{-9}$; the short-range $\operatorname{erfc}$ correction remains a screened direct sum. The point-charge term is recomputed directly within 0.01~Bohr of an atom to avoid cancellation error in the two-term representation.

Both AD and Hybrid use the same Gaussian density and gradient. Each shell receives a spatial radius beyond which the absolute density and gradient tails are each below $10^{-9}$ in atomic units. AD additionally screens the short-range Coulomb correction where its absolute tail is below $10^{-9}$. These are per-shell cutoffs, not a bound on the sum over all shells. Hybrid instead evaluates the global Gaussian Coulomb matrix analytically from three-centre integrals; it uses neither the AD FMM nor the short-range Coulomb sum.

Both methods also use the same learned local field $\Phi_\theta$, which vanishes outside the 5~\text{\AA} atom--grid cutoff $R_c$. Large-system AD evaluates this local field using smoothed atom--grid distances $r_\epsilon=\sqrt{r^2+\epsilon^2}$ with $\epsilon=0.03$~Bohr to stabilize higher-order spatial derivatives. AD differentiates this field to obtain the local density and its gradient, whereas Hybrid uses the seven-point stencil for the density and first-order AD for its gradient. AD constructs local atom--grid edges within $R_c$. For Hybrid at $h=0.03$~Bohr, candidate atoms are found once within $R_c+h+10^{-5}$~Bohr of each central point. The exact $r<R_c$ cutoff is then applied at every displaced stencil point, so the larger search radius does not change the field's support.

\end{document}